\documentclass[a4paper,10pt,accepted=2022-03-15]{quantumarticle}
\pdfoutput=1
\usepackage[utf8]{inputenc}
\usepackage[T1]{fontenc}
\usepackage[english]{babel}
\usepackage{amsmath}
\usepackage{amssymb}
\usepackage{amsfonts}
\usepackage{mathrsfs}
\usepackage{bm, dsfont}
\usepackage{booktabs}
\usepackage{multirow}
\usepackage[usenames,dvipsnames]{color}
\usepackage{graphicx}
\usepackage[backend=bibtex,sorting=none,style=trad-abbrv,citestyle=numeric-comp]{biblatex}
\usepackage[colorlinks=true,linkcolor=blue,citecolor=blue,urlcolor=blue]{hyperref}
\usepackage{cleveref}
\usepackage{hypcap}
\usepackage{siunitx}

\newcommand{\bra}[1]{{\langle{#1}|}}
\newcommand{\ket}[1]{{|{#1}\rangle}}
\newcommand{\ketbra}[2]{{|{#1}\rangle\!\langle{#2}|}}

\newcommand{\ev}[1]{\langle{#1}\rangle}
\newcommand{\tr}{\text{tr}}

\newcommand{\ii}{\mathrm{i}}
\newcommand{\id}{\mathds{1}}

\newcolumntype{C}{>{$}c<{$}}
\AtBeginDocument{
\heavyrulewidth=.08em
\lightrulewidth=.05em
\cmidrulewidth=.03em
\belowrulesep=.65ex
\belowbottomsep=0pt
\aboverulesep=.4ex
\abovetopsep=0pt
\cmidrulesep=\doublerulesep
\cmidrulekern=.5em
\defaultaddspace=.5em
}

\begin{document}
\title{Local and scalable detection of genuine multipartite single-photon path entanglement}

\author{Patrik~Caspar}
\affiliation{Department of Applied Physics, University of Geneva, CH-1211 Genève, Switzerland}
\author{Enky~Oudot}
\affiliation{ICFO - Institut de Ciencies Fotoniques, The Barcelona Institute of Science and Technology, 08860 Castelldefels (Barcelona), Spain}
\author{Pavel~Sekatski}
\affiliation{Department of Applied Physics, University of Geneva, CH-1211 Genève, Switzerland}
\author{Nicolas~Maring}
\affiliation{Department of Applied Physics, University of Geneva, CH-1211 Genève, Switzerland}
\author{Anthony~Martin}
\affiliation{Department of Applied Physics, University of Geneva, CH-1211 Genève, Switzerland}
\thanks{Current address: Université Côte d'Azur, CNRS, Institut de Physique de Nice, Parc Valrose, F-06108 Nice Cedex~2, France}
\author{Nicolas~Sangouard}
\affiliation{Institut de physique théorique, Université Paris Saclay, CEA, CNRS, F-91191 Gif-sur-Yvette, France}
\author{Hugo~Zbinden}
\affiliation{Department of Applied Physics, University of Geneva, CH-1211 Genève, Switzerland}
\author{Rob~Thew}
\email{Robert.Thew@unige.ch}
\affiliation{Department of Applied Physics, University of Geneva, CH-1211 Genève, Switzerland}
\begin{abstract}
How can a multipartite single-photon path-entangled state be certified efficiently by means of local measurements? We address this question by constructing an entanglement witness based on local photon detections preceded by displacement operations to reveal genuine multipartite entanglement. Our witness is defined as a sum of three observables that can be measured locally and assessed with two measurement settings for any number of parties $N$. For any bipartition, the maximum mean value of the witness observable over biseparable states is bounded by the maximum eigenvalue of an $N\times N$ matrix, which can be computed efficiently. We demonstrate the applicability of our scheme by experimentally testing the witness for heralded 4- and 8\nobreakdash-partite single-photon path-entangled states. Our implementation shows the scalability of our witness and opens the door for distributing photonic multipartite entanglement in quantum networks at high rates. 
\end{abstract}

\maketitle

\section{Introduction}
The generation, distribution and certification of entanglement in multipartite quantum communication networks is of increasing importance as the size and complexity of networks grow beyond simple short-distance point-to-point scenarios~\cite{Kimble2008, Wehner2018}. 
In general, multipartite entanglement enables applications such as enhanced sensing~\cite{Komar2014, Liu2021} or multi-user quantum communication protocols~\cite{Murta2020, Lipinska2018}.
At the heart of the matter, the challenge is to find scalable solutions to realize these applications, which still remain experimentally feasible.
On the one hand, the experimental limitations of probabilistic multi-photon sources, especially in terms of rates~\cite{Zhong2018}, represent fundamental obstacles for entangled state generation. On the other hand, for the state certification, the exponential scaling of measurements in tomography as the number of parties increases~\cite{James2001} makes it impractical already for a small number of parties.

\begin{figure}[b]
\capstart
\includegraphics[width = \columnwidth]{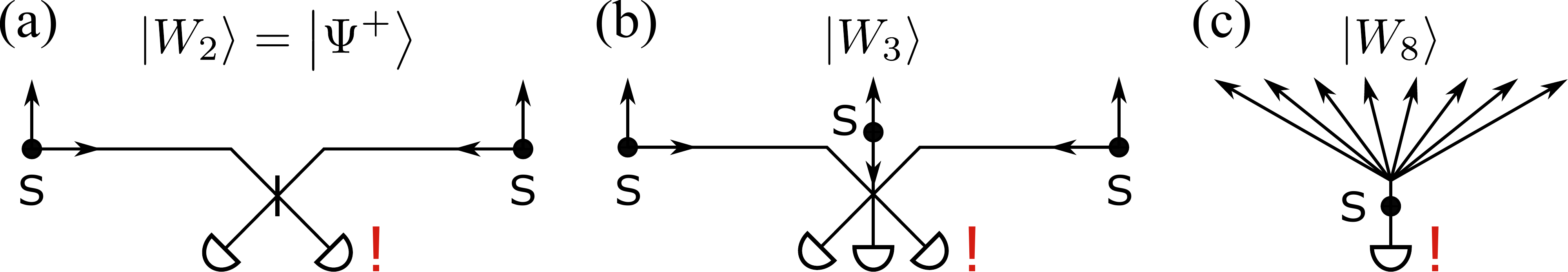}
\caption{\label{fig:Concept} Heralded multipartite entanglement distribution. (a) Entanglement is distributed between two parties, each having a signal-idler photon pair source (S). The idler modes are combined on a beam splitter and the detection of a single photon after this beam splitter projects the signal modes into a single-photon entangled state. (b) Generalization of the scheme to tripartite states while keeping the local losses low. The aim of this work is to clarify on how entanglement can be detected in this setting. (c) Conceptual schematic of the experiment. Entanglement is distributed among several parties by locally splitting the signal mode into multiple spatial output modes.}
\end{figure}

In the case of generation and distribution of entanglement already for quantum repeaters, a shift away from photon-pair to heralded single-photon entanglement provided significant scaling benefits even for point-to-point communication schemes~\cite{Sangouard2011}. For example, as shown in Fig.~\hyperref[fig:Concept]{\ref{fig:Concept}(a)}, the distribution of entanglement between two remote parties through optical fiber can be realized efficiently by giving each party a source emitting signal-idler photon pairs, combining the idler modes into a beam splitter at a central station and placing two detectors at the output of the beam splitter. A photon detection by one of the two detectors heralds the sharing of a single photon between the signal modes -- a single-photon path-entangled state~\cite{Duan2001,Simon2007}. Although the realization of such schemes faces the challenge of active stabilization of the phase between the two parties \cite{Minar2008}, work addressing this issue has been reported \cite{Chou2005,Slodi2013,Delteil2016,Stockill2017,Humphreys2018,Lago-Rivera2021}, even over longer distances in optical fiber \cite{Caspar2020,Yu2020}. 

More interestingly, this approach can be efficiently extended to the distribution of entanglement between multiple parties by simply replacing the two-port beam splitter by a multi-port beam splitter, see Fig.~\hyperref[fig:Concept]{\ref{fig:Concept}(b)}. This represents an efficient way of generating a multipartite entangled state close to a W state~\cite{Duer2000}:
\begin{equation}\label{eq:state}
\ket{W_N} = \frac{1}{\sqrt{N}}\sum_{i=1}^N \ket{0,...,0,1_i,0,...,0}
\end{equation}
with a single photon detection heralding its successful distribution remotely. Here, $\ket{0}$ denotes the vacuum state, $\ket{1}$ the single-photon number state and $N$ the number of parties. Applications of this class of distributed states include long-baseline telescopes that can take advantage not only of a distribution over long distances but also the multipartite setting~\cite{Gottesman2012,Khabiboulline2019a,Khabiboulline2019b}.

The certification of entanglement in such multipartite quantum communication networks is extremely challenging. Beyond tomography, even typical entanglement witnesses require multiple settings per party~\cite{Morin2013,Monteiro2015}, with the commensurate scaling that quickly ensures their infeasibility. Others have used certification techniques that require the recombination of optical modes~\cite{Papp2009,Graefe2014}, which are impractical in communication scenarios where local measurements are required. These problems are further exacerbated in a distributed setting, where noisy and lossy channels also need to be addressed.

Here we develop an entanglement witness tailored to the W state, given in Eq.~\eqref{eq:state}, to reveal genuine multipartite entanglement (GME) without postselection. We assume that the measurement apparatus is well characterized and that each party $i$ holds a single optical mode with an associated bosonic annihilation operator $a_i$. There are no further assumptions. In particular, the photon number statistics are unknown and we do not assume that the same state is prepared in each run (i.i.d.). The witness is scalable as it only requires two different measurement settings, independently of the number of parties. Its applicability is demonstrated using an experimental setup in a configuration like in Fig.~\hyperref[fig:Concept]{\ref{fig:Concept}(c)} where genuine 4- and 8\nobreakdash-partite entangled states are heralded at high rates and successfully verified. 

\section{Theory}
To certify the GME of $N$-partite single-photon path-entangled states, we build a witness using practical single-photon detectors, i.e. non-unit efficiency and non-photon number resolving. Such a detector can be modeled as a loss channel with transmission $\eta$ (the detector efficiency) followed by a two-outcome measurement that perfectly distinguishes the vacuum $\ket{0}$ from all the other Fock states of the detected mode. In this model, the fixed loss can be interpreted as part of the state preparation degrading its entanglement. In the following, we therefore model the detector operating on party $i$ with the positive operator valued measure (POVM) $\{E_c^i= \mathds{1} - \Pi^{(i)}_0, E_0^i=\Pi^{(i)}_0\}$, $E_c^i$ ($E_0^i$) corresponding to the POVM element associated to a click $(c)$ (no-click $(0)$) event and $\Pi^{(i)}_0=\ketbra{0}{0}$ is the projection on the vacuum. With such a detector, we thus have access to the weight of the vacuum component for each party from the no-click events, i.e. from $\tr (E_0^i\,\rho)$ where $\rho$ denotes the $N$-mode state produced in the actual experiment. By probing each mode with such a detector, we can access the probability that more than one mode contain photons $\sum_{n\geq 2} P_\text{click}^n$, where $P_\text{click}^n$ is the probability that $n$ detectors click. Let us define the POVM elements associated with these probabilities as $E_{n\geq 2}$ and $E_n$. Note that $p_0 = P_\text{click}^0 = \tr(\ketbra{0}{0}^{\otimes N} \,\rho)$ is the probability that the state contains no photons. 
Furthermore, by probing the mode $i$ with two such detectors after a 50/50 beam splitter, we can upper bound the probability that it contains more than one photon~\cite{Monteiro2015}, that is $\tr (\Pi^{(i)}_{n_i\geq 2}\,\rho )$ with $\Pi^{(i)}_{n_i\geq 2}=\sum_{n_i\geq2}\ketbra{n_i}{n_i}$. This allows us to upper bound the probability that the $N$-mode state $\rho$ contains two or more photons, which we write as an operator inequality
\begin{equation}\label{eq:p*}
\Pi_{n\geq 2} \leq E_{n\geq 2} + \sum_{i} \Pi^{(i)}_{n_i\geq 2},
\end{equation}
where $\Pi_{n\geq 2}$ is the projector on all the combinations of Fock states containing at least two photons in total. We then denote $p_* =\tr((E_{n\geq 2} + \sum_{i} \Pi^{(i)}_{n_i\geq 2})\,\rho)$.

In order to implement measurements that are sensitive to the coherence between different products of Fock states, the party $i$ can perform a phase-space displacement operation $D(\alpha_i)=e^{\alpha_i a_i^\dag - \alpha_i^* a_i}$ right before the detector~\cite{Paris1996}. As shown in Appendix~\ref{sec:app_deteff}, the loss after the displacement operation, i.e. the detector inefficiency, can be permuted with the displacement by only adjusting the displacement amplitude $\alpha_i$. This allows us to keep the POVM above and define a parametric family of local observables for each party $i\in\{1,...,N\}$ ~\cite{Vivoli2015}
\begin{equation}
\begin{split}
    \sigma_{\alpha_i}^{(i)} &= D^\dag(\alpha_i)(E_0^i- E_c^i) D(\alpha_i) \\
    &= D^\dag(\alpha_i)(2\ketbra{0}{0}- \mathds{1}) D(\alpha_i)
\end{split}
\end{equation}
by attributing the value $+1$ to a no-click and $-1$ to a click event. In principle, asking each party to perform such a measurement and combining the results allows us to define a global observable \begin{equation}
\hat{O}_{\bm \alpha}=\sum^N_{i\neq j}\sigma_{\alpha_i}^{(i)}\otimes\sigma_{\alpha_j}^{(j)},
\end{equation}
where $\bm \alpha = (\alpha_1,\dots, \alpha_N)$. To simplify the experimental realization, we consider the case where the local oscillators used by each party to perform the displacement operations are not phase-locked to the input state, so that the local displacements $\alpha_i\mapsto \alpha_i e^{\ii \varphi}$ are only defined up to an arbitrary common phase $\varphi$, but the phase differences between the parties are well controlled and kept constant at zero. As a consequence, we measure the observable
\begin{equation}\label{eq: twirling}
  \hat{\mathcal{O}}_{\bm \alpha} = \frac{1}{2\pi} \int_{0}^{2\pi} \textrm{d}\varphi\ e^{-\ii \varphi \sum_{i=1}^N a_i^\dag a_i}\ \hat{O}_{\bm \alpha}\ e^{\ii \varphi \sum_{i=1}^N a_i^\dag a_i}.  
\end{equation}
Because of the phase averaging, the operator $\hat{\mathcal{O}}_{\bm \alpha} =\bigoplus_{n=0}^\infty \hat{\mathcal{O}}_{\bm \alpha}^{(n)}$ acts orthogonally on different total photon number subspaces $n =\ev{\sum_{i=1}^N a_i^\dag a_i}$.

For displacement amplitudes chosen in the appropriate range~\cite{Vivoli2015}, each two-body correlator $\sigma_{\alpha_i}^{(i)}\otimes\sigma_{\alpha_j}^{(j)}$ gathers the coherence $\ketbra{1_i,0_j}{0_i, 1_j}+h.c.$. This coherence is symptomatic of GME in the state $\ket{W_N}$, making the observable $\hat{\mathcal{O}}_{\bm \alpha}$ a natural candidate to witness this entanglement. However, $\hat{\mathcal{O}}_{\bm \alpha}$ is also sensitive to all higher photon number contributions, which are difficult to characterize. A simple way to circumvent this problem is to subtract the observable $N(N-1) \Pi_{n\geq 2}$ from $\hat{\mathcal{O}}_{\bm \alpha}$. On the one hand, for our state $\tr (\Pi_{n\geq 2}\, \rho) \approx 0$ and the expected value is not affected much by the subtraction. On the other hand, since $\|\hat{\mathcal{O}}_{\bm \alpha}\|\leq N(N-1)$, one can bound
\begin{equation} \label{eq: W up}
\hat{\mathcal{O}}_{\bm \alpha} - N(N-1) \Pi_{n\geq 2} \leq \hat{\mathcal{O}}_{\bm \alpha}^{(0)}\oplus\hat{\mathcal{O}}_{\bm \alpha}^{(1)}
\end{equation}
by an operator supported on a subspace with at most one photon.

We can now define the entanglement witness
\begin{equation}\label{eq:witness}
  \hat{\mathcal{W}}_{\bm \alpha}= \hat{\mathcal{O}}_{\bm \alpha}    -  N(N-1) \Pi_{n\geq 2} + M_{n\leq 1} - \mu E_{n\geq 2},
\end{equation}
where $M_{n\leq 1}$ is an operator diagonal in the product Fock basis and acting on the subspace with one photon at most, specified in Appendix~\ref{sec:app_witness}, and $\mu$ is a positive real parameter that one can tune. To show that the witness can reveal GME, let us start by computing the biseparable bound.
\begin{equation}
\label{eq_maxrhobisep}
    w_{\text{bisep}} = \max_{\varrho_\textrm{bisep}} \tr ( \hat{\mathcal{W}}_{\bm \alpha} \varrho_\textrm{bisep} ),
\end{equation}
i.e. the maximum value the witness takes on any biseparable state. A general biseparable state is a mixture of states that are product states for some bipartition (a partition of all modes into two groups). Formally,
\begin{equation}\label{def: bisep}
\varrho_\textrm{bisep} = \sum_{G_1|G_2} p(G_1|G_2) \, \rho_{G_1|G_2}.
\end{equation}
Here, the sum runs over all partitions $G_1|G_2$ of the $N$ parties where $G_1\cup G_2=\{1,2,\dots,N\}$ and $G_1\cap G_2=\emptyset$. The probabilities of different partitions sum up to one $\sum_{G_1|G_2} p(G_1|G_2)=1$ and $\rho_{G_1|G_2}$ is a separable state with respect to the partition $G_1|G_2$. Since the set of biseparable states is convex, the maximum value that an observable takes on any biseparable state $\varrho_\textrm{bisep}$, including mixed states, 
\begin{equation}\label{eq: w bisep}
w_\text{bisep}
=\max_{G_1,G_2,\ket{\Psi}} \bra{\Psi}\hat{\mathcal{W}}_{\bm \alpha}\ket{\Psi},
\end{equation}
is attained for a pure state $\ket{\Psi}= \ket{\Psi_1}_{G_1}\ket{\Psi_2}_{G_2}$ on some partition. Now, using Ineq.~\eqref{eq: W up} we obtain a relaxation
\begin{align}
    w_\text{bisep} &\leq \max_{G_1,G_2,\ket{\Psi}} \bra{\Psi}  \widetilde{\mathcal{W}} \ket{\Psi},\\
    \widetilde{\mathcal{W}} & = \hat{\mathcal{O}}_{\bm \alpha}^{(0)}\oplus\hat{\mathcal{O}}_{\bm \alpha}^{(1)} + M_{n\leq 1} - \mu E_{n\geq 2}
\end{align}
which simplifies the maximization problem enormously. Indeed, the operator $\widetilde{\mathcal{W}}$ is block diagonal, with restriction to the sector with two or more photons $-\mu E_{n\geq 2}$ that is negative. Thus, we can restrict the maximization to states $\ket{\Psi_k}_{G_k}$ which contain one photon at most and write
\begin{equation}
\ket{\Psi_k}_{G_k} = v_0^{(k)} \ket{0}_{G_k} + \sum_{i=1|j_i\in G_k}^{|G_k|} v_{i}^{(k)} a^\dag_{j_i}\ket{0}_{G_k}.
\end{equation}
For a fixed partition, the product states can be parametrized by normalized vectors $\bm v^{(1)}$ and $\bm v^{(2)}$, that can be taken to be real without loss of generality. This leaves us with a $N$-parameter optimization problem. In Appendix~\ref{sec:app_witness}, we show that for our witness this maximization can be reduced to a single parameter optimization
\begin{equation}
\begin{split}
 \widetilde{w}_{G_1,G_2} &=  \max_{\bm v^{(1)},\bm v^{({2})}} \bra{\Psi} \widetilde{\mathcal{W}}  \ket{\Psi} \\
 &= \max_{a \in [0,2\pi]} (\text{max\,eig}(\mathds{M}(\lambda, \mu, \bm \alpha, a)))
\end{split}
\end{equation}
of the maximum eigenvalue of a $N\times N$ matrix $\mathds{M}(\lambda, \mu, \bm \alpha, a)$, which can be solved efficiently with standard numerical tools. Here, $\lambda$ (like $\mu$) is a positive real parameter of the witness that can be tuned. We solve the optimization for all bipartitions to obtain a relaxation of the biseparable bound for the witness $\hat{\mathcal{W}}$
\begin{equation}
w_\text{bisep}\leq \widetilde{w}_\text{bisep} = \max \limits_{G_1,G_2} (\widetilde{w}_{G_1,G_2}).
\end{equation}

It remains to explain how we estimate the violation of the witness $\langle \hat{\mathcal{W}}_{\bm \alpha} - \widetilde{w}_\text{bisep} \rangle$ on the multimode state $\rho$ prepared in the experiment. In reality, the amplitudes $\bm \alpha$ of the displacements fluctuate within some range $\bm \alpha \in A$ that we characterize. The operator $\hat{\mathcal{W}}_{\bm \alpha}$ depends on these amplitudes both "physically" via the observable $\hat{\mathcal{O}}_{\bm \alpha}$, but also "algebraically" via the definition of the operator $M_{n\leq 1}(\bm \alpha)$ and the value for the biseparable bound $\widetilde{w}_\text{bisep}(\bm \alpha)$. To remove the second dependence, we consider the worst-case scenario $w_\text{bisep}^\textrm{max} = \max_{\bm \alpha \in A} \widetilde{ w}_\text{bisep}(\bm \alpha)$ and $\bar{M}_{n\leq 1} = \min_{\bm \alpha \in A}  M_{n\leq 1}(\bm \alpha)$. Then, defining the operator $\overline{\mathcal{W}}_{\bm \alpha}$, where we replace $M_{n\leq 1}$ by $\bar{M}_{n\leq 1}$ in the witness (Eq.~\eqref{eq:witness}), implies
\begin{equation}
\hat{\mathcal{W}}_{\bm \alpha} - w_\text{bisep}   \geq  \overline{\mathcal{W}}_{\bm \alpha} - w_\text{bisep}^\textrm{max}.
\end{equation}

\begin{figure*}
\capstart
\includegraphics[width = \textwidth]{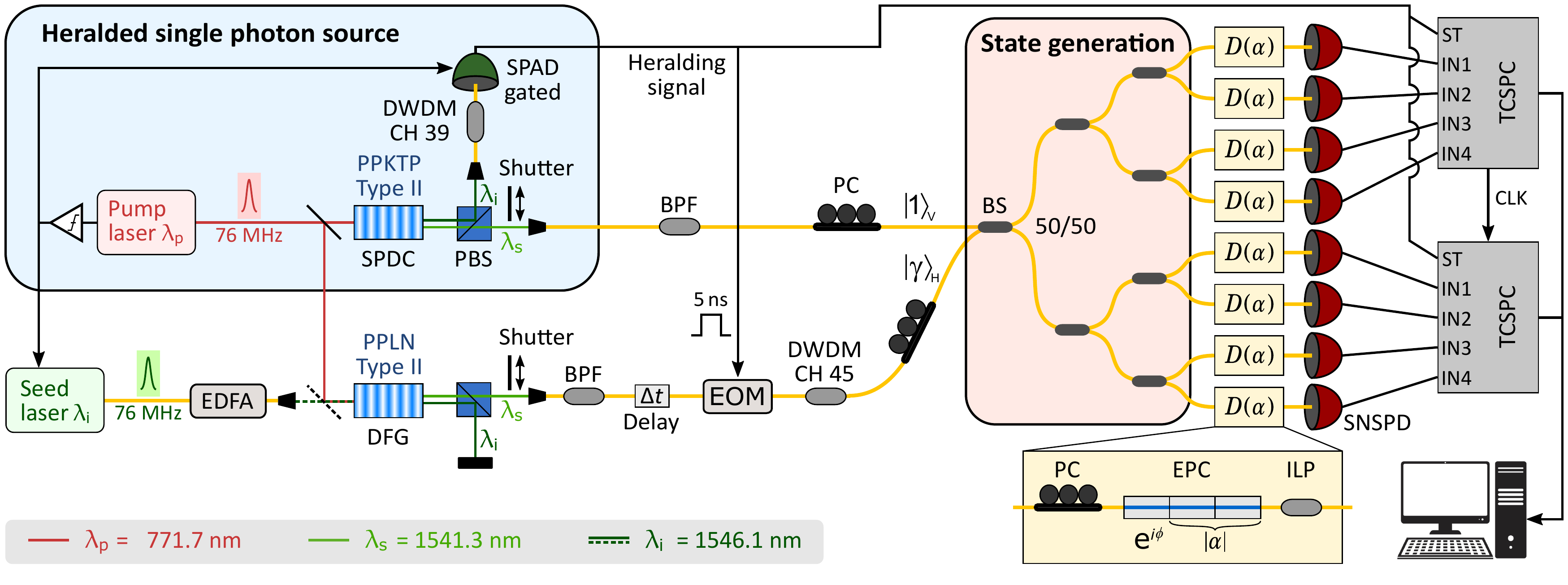}
\caption{\label{fig:setup} Experimental schematic: A heralded single photon, incident on a cascade of 50/50 fiber beam splitters (BS), is delocalized over spatial modes to generate an 8\nobreakdash-partite path-entangled state. Weak coherent states in orthogonal polarization modes are co-propagated with the single-photon state to locally perform displacement-based measurements. See main text for details on the setup and notation.}
\end{figure*} 

To prove GME it suffices to show that the average (over all rounds with fluctuating $\bm \alpha$) expected value of $\langle \overline{\mathcal{W}_{\bm \alpha}} \rangle$ on $\rho$, which we call $w_\rho^\text{max}$, exceeds the constant $w_\text{bisep}^\textrm{max}$. As argued in Appendix~\ref{sec:app_measuring}, $\langle \overline{\mathcal{W}_{\bm \alpha}} \rangle$ can be estimated by combining the average values of three different observables measured independently in different runs of the experiment. These are, $\hat{\mathcal{O}}_{\bm \alpha}$ measured with displacement operations, $\mathcal{Z}$ as defined in Eq.~\eqref{eq:W_app} and measured without displacing and one detector per mode, and $\Sigma_{n\geq 2}= \sum_{i} \Pi^{(i)}_{n_i\geq 2}$ measured on a single mode with a 50/50 beam splitter and two detectors. Finally, in Appendix~\ref{sec:app_statistics} we analyze the statistical significance of the observed violation of the witness. We use Hoeffding's theorem (1963) \cite{Hoeffding1963} to upper bound the $p$-value for the null-hypothesis that the state $\rho$ is biseparable.

\section{Experiment}
The experimental setup is presented in Fig.~\ref{fig:setup}. We use a heralded single photon source (HSPS) employing a periodically poled potassium titanyl phosphate (PPKTP) nonlinear crystal as a type-II spontaneous parametric down-conversion (SPDC) source. The crystal is pumped by a Ti:Sapphire laser at $\lambda_p=\SI{771.7}{nm}$ in the picosecond pulsed regime with a repetition rate of \SI{76}{MHz} to create nondegenerate photon pairs at $\lambda_s=\SI{1541.3}{nm}$ (signal) and $\lambda_i=\SI{1546.1}{nm}$ (idler). The pair creation probability per pump pulse is kept at $p_\mathrm{pair} \approx \num{2.7E-3}$ in order to minimize the impact of double-pair emissions. Signal and idler modes are separated after their generation by a polarizing beam splitter (PBS) and coupled into single-mode fibers (SMF). By spectrally filtering the heralding idler photon using a dense wavelength division multiplexer (DWDM) we ensure high-purity heralded signal photons.

Detection of one heralding photon by an InGaAs single-photon avalanche diode (SPAD -- ID Quantique ID210) in gated mode with a detection efficiency of around \SI{20}{\%} heralds the presence of a fiber-coupled signal photon with a heralding efficiency of around $\SI{75}{\%}$.
The heralded signal photon first encounters a band-pass filter (BPF) with a passband between \SI{1528}{nm} and \SI{1565}{nm} in order to further remove residual pump light and is subsequently sent to a cascade of 50/50 fiber beam splitters (BS) where it is delocalized to generate the targeted multipartite path-entangled state.
In this manner, we herald the entangled state at a rate of \SI{11.5}{kcps} where \SI{0.6}{kcps} are attributed to dark counts, which effectively adds loss to the state.
 
To generate the coherent state with the same spectral and temporal properties as the signal photon for the displacement-based measurement (see Appendix~\ref{sec:app_experiment}), we stimulate a difference frequency generation (DFG) process in a type-II quasi phase-matched periodically poled lithium niobate (PPLN) nonlinear crystal~\cite{Bruno2014}. To this end, the crystal is pumped by the same laser pulses as the HSPS and seeded with pulses at the same repetition rate originating from a distributed feedback (DFB) laser at $\lambda_i=\SI{1546.1}{nm}$.
The seed laser is driven from well below to above the lasing threshold each cycle to phase randomize the coherent state and in order to reach the required displacement amplitude, we amplify the pulses with an Erbium doped fiber amplifier (EDFA). 
We then couple the coherent state into SMF and further filter out residual pump light with a BPF before adjusting the time delay between the coherent state and the signal photon with a motorized delay line. Moreover, to avoid saturation of the detectors (see below), we select the coherent state pulses by passing them through an electro-optic (amplitude) modulator (EOM) with an extinction ratio of $\sim\SI{30}{dB}$ triggered by a \SI{5}{ns} gate upon successful detection of a heralding photon. Residual seed laser light is filtered with a DWDM at $\lambda_s$. The coherent state is then sent into the second port of the first 50/50 BS where fiber polarization controllers (PC) are used to ensure orthogonal polarizations between coherent and single-photon states. The subsequent co-propagation passively guarantees stability of the relative optical phase between the coherent and single-photon states.

In order to perform the local displacement operations in an all-fiber configuration, the coherent and single-photon states are projected onto the same polarization mode using an in-line polarizer (ILP) preceded by a manual PC and a three-segment electronic polarization controller (EPC -- Phoenix Photonics PSC). 
The first segment of the EPC allows for the control of the relative phase between the coherent and single-photon states (see Appendix~\ref{sec:app_experiment}), whereas the second and third segments are used to control the displacement amplitude. In each spatial mode, the coherent state is set to have a mean photon number per pulse of roughly 13 before the polarizer to achieve a displacement amplitude of $\alpha\approx 0.83$. Drifts and fluctuations in the displacement amplitudes during the data acquisition are taken into account for the evaluation of the witness (see Appendix~\ref{sec:app_experiment}). 
The photons are detected by eight in-house-developed MoSi superconducting nanowire single-photon detectors (SNSPD) with detection efficiencies between $\SI{75}-\SI{82}{\%}$~\cite{Caloz2018}. Time-correlated single-photon counting (TCSPC) using two clock-synchronized programmable time-to-digital converters (ID Quantique ID900) is then used to register detections conditioned on a successful heralding event. 

After the alignment of the relative phases between the output modes, data for the witness is acquired in 20 sequences of \SI{5}{min}. Each sequence measures the displacement amplitudes (\SI{1}{min}) and then $\rho$ with and without displacement (\SI{2}{min} each). Shutters in the corresponding paths (see Fig.~\ref{fig:setup}) are used to switch between the measurements. In order to estimate the probability of having more than one photon locally, we additionally perform a heralded autocorrelation measurement on one output mode by inserting a 50/50 BS before the detectors and acquire data for \SI{6}{h}.

\section{Results}

\begin{table}
\capstart
\centering
\small
\setlength\tabcolsep{3.5pt}
\begin{tabular}{c c c c c c} 
\toprule
$N$ & $p_0$ & $p_*$ & $w_\rho^\mathrm{exp}$ & $w_\mathrm{bisep}^\mathrm{max}$ & $p$-value \\[0.5ex]
\cmidrule{1-6}
4 & 0.6891(1) & $4.36(5)\cdot 10^{-4}$ & 2.993(2) & 2.785 & $10^{-1952}$ \\
8 & 0.7319(1) & $3.40(4)\cdot 10^{-4}$ & 8.565(4) & 8.358 & $10^{-87}$ \\
\bottomrule
\end{tabular}
\caption{\label{tab:ResultsWitness} Measured expectation values $w_\rho^\mathrm{exp}$ and calculated separable bounds $w_\mathrm{bisep}^\mathrm{max}$ of the $N$-partite entanglement witness for states with vacuum contributions $p_0$ and upper bounds $p_*$ on the probability of having more than one photon, see Ineq.~\eqref{eq:p*}. The $p$-value for the null-hypothesis that the state $\rho$ is biseparable is calculated according to Appendix~\ref{sec:app_statistics}.}
\end{table}

\begin{figure}
\capstart
\includegraphics[width = 0.96\columnwidth]{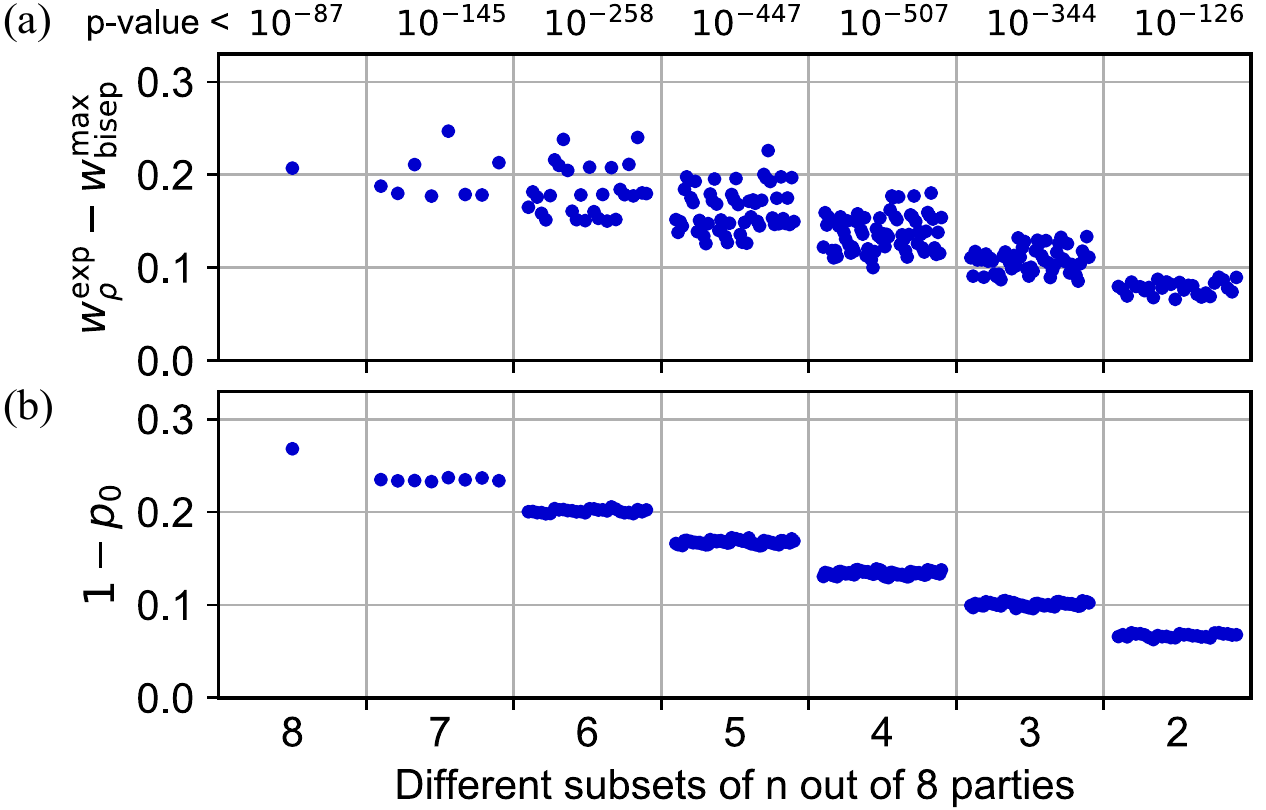}
\caption{\label{fig:witness} Results of the witness for $N=8$. The witness is applied to all possible different subsets of $n$ out of $N$ parties where the discarded parties are traced out. The plots show (a) the violation of the witness $w_\rho^\mathrm{exp}-w_\mathrm{bisep}^\mathrm{max}$ with the maximum $p$-value for each fixed number of subsets indicated on top and (b) $1-p_0$ where $p_0$ is the probability of the vacuum component for each subset of parties.}
\end{figure}

The witness is measured for two different experimental configurations with $N=4$ and 8 parties and the separable bound is violated in both cases, as shown in Tab.~\ref{tab:ResultsWitness}. In the case of $N=8$ we have more loss on the state, mainly due to the insertion loss of another BS and the lower detection efficiencies of the four additional SNSPDs. 

For $N=8$ we further analyze the data by considering all $\sum_{n=2}^N \genfrac(){0pt}{2}{N}{n} = 247$ possible subsets of $n\in\{2,...,N\}$ out of $N$ parties and calculating for each subset the expectation value and separable bound of the $n$-partite witness. The results are presented in Fig.~\ref{fig:witness}. It is expected for our state that all subsets of parties show GME, however, these results suggest that also for a high probability of having vacuum for all parties $p_0$, our witness is suitable to detect GME. We attribute the fact that the witness violation $w_\rho^\mathrm{exp}-w_\mathrm{bisep}^\mathrm{max}$ varies for different choices of the same number of parties $n$ to the difference in transmission and detection efficiencies for different parties (see Appendix~\ref{sec:app_experiment}).

\section{Discussion}
Let us discuss the scalability of our witness with the number of parties $N$. First, we emphasize that our witness only relies on three measured quantities. Two of them, the measurement of $p_0$ and $p_*$, can be obtained with a single setting per party corresponding to no displacement. The last quantity is assessed with a second setting using displacement operations. This is in contrast with other methods where the overall number of settings grows polynomially with $N$, and in sharp contrast with techniques relying on state tomography where it grows exponentially with $N$. The second aspect is the computational resources required to compute the biseparable bound. In our case, we only need to compute the maximum eigenvalue of an $N\times N$ matrix for each bipartition of the $N$ parties in two groups. For any bipartition, the computational complexity of constructing the matrix and computing its maximum eigenvalue scales much better than the methods relying on a relaxation of a semi-definite optimization over the biseparable states of $N$ qubits proposed earlier, e.g.~\cite{Monteiro2015}. Finally, the number of bipartitions to check grows exponentially $2^{N-1}-1$ if all the amplitudes of displacement operations are different, however, this is reduced to $\lfloor N/2 \rfloor$ if all the displacement amplitudes can be assumed to be equal $\alpha_i = \alpha_j$. In our experimental implementation, this could be achieved by actively stabilizing the power of the coherent state before inserting it into the first beamsplitter. Further, active control of the EPCs as part of the local measurement setup would suppress drifts in the displacement amplitudes.
\medskip

\begin{figure}[tb]
\capstart
\includegraphics[width = 0.94\columnwidth]{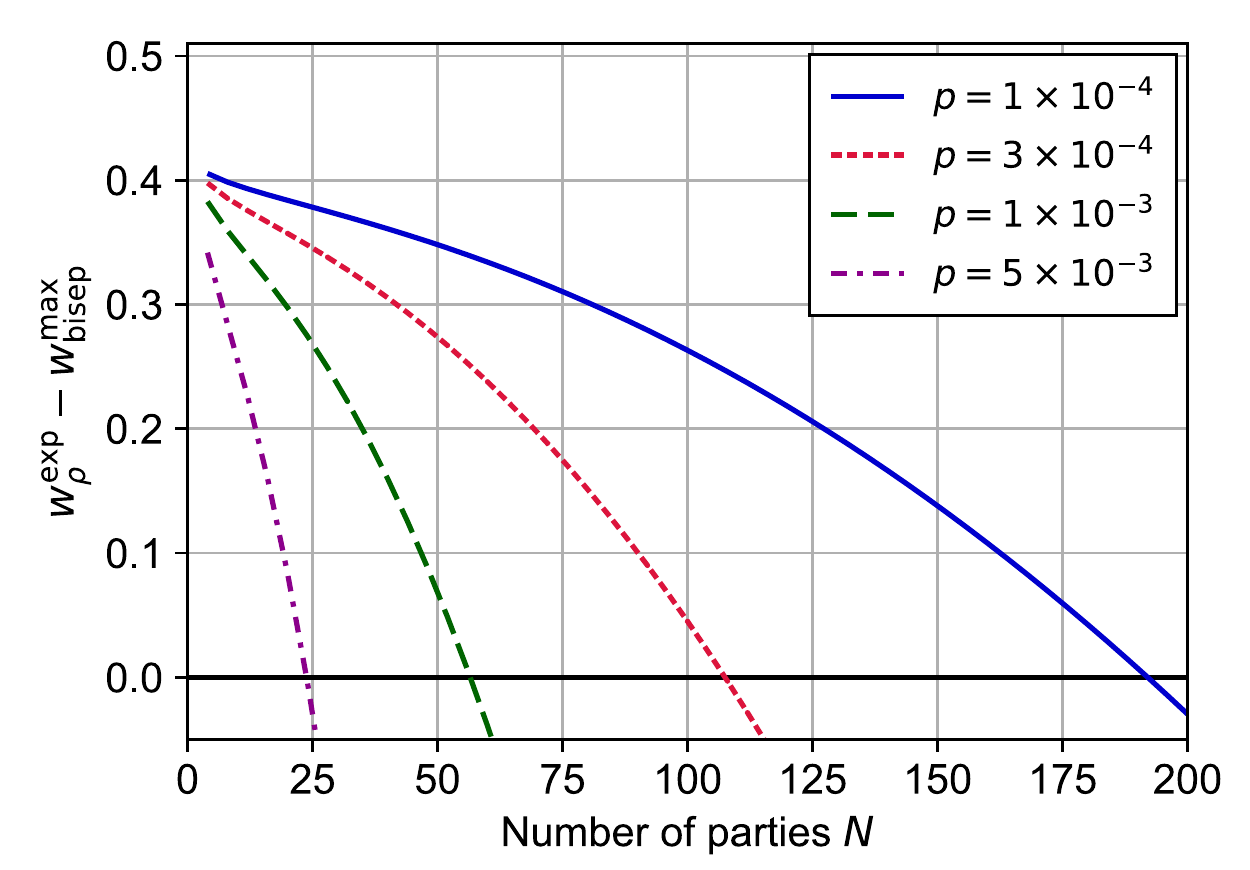}
\caption{\label{fig:scalability} Calculation of the witness violation $w_\rho^\mathrm{exp}-w_\mathrm{bisep}^\mathrm{max}$ as a function of the number of parties $N$ for a state $\rho=(\ketbra{1}{1}+p\ketbra{2}{2})/(1+p)$ input to a $N$-port beam splitter after undergoing a loss channel with transmission $\eta=0.3$. The two-photon probability $p=\num{5E-3}$ is a good approximation for the state generated in the experiment. In this calculation, we assume a perfectly balanced state and fix the displacement amplitude for the measurement to $\alpha=\sqrt{\ln{2}}\approx 0.83$ for each party, which is the most robust to fluctuations in $\alpha$.}
\end{figure}

In the experiment, the main limiting factor for the demonstration of GME in large systems using the presented witness are contributions to the state with more than one photon in total, which we upper bound by $p_*$. In order to estimate the maximum number of parties $N$ for which the witness still applies, we calculate the expected witness violation for a state $\rho=(\ketbra{1}{1}+p\ketbra{2}{2})/(1+p)$ that undergoes a loss channel with transmission $\eta=0.3$ and is then equally split into $N$ modes, which is a good approximation of the state created in the experiment. 
As shown in Fig.~\ref{fig:scalability}, we see that the reduction of the probability of generating a two-photon state increases the number of parties for which our witness is able to detect GME. For a state similar to the one in the experiment, this would allow for the demonstration of GME for up to 23 parties. Furthermore, we show in Appendix~\ref{sec:app_dark_counts_GME} that our witness can be directly used in the presence of dark counts if one adds a term to the biseparable bound. We check that this does not affect the demonstration of GME in our experiment and further investigate the scalability under the presence of dark counts.

\section{Conclusion}
The developed witness is well suited for efficient certification of multipartite single-photon path entanglement in future quantum networks. Highly entangled multipartite states could be distributed at high rates in a scheme where each party holds a photon-pair source and one photon of each pair is sent to a central multi-port beam splitter that erases the which-path information. In this way, local losses can be kept low and the added distance between parties only reduces the heralding rate. In combination with quantum memories, such a scheme has potential for applications relying on distributed W states. The experimental challenge in such a scheme, however, remains the need for phase stability in long fiber links. 

\begin{acknowledgments}
The authors would like to thank F.~Bussières, M.~Caloz and M.~Perrenoud for the development of the SNSPDs and C.~Barreiro for technical support. This work was supported by the Swiss National Science Foundation SNSF (Grant No.~200020\_182664), the NCCR QSIT and by the European Union’s Horizon 2020 research and innovation program under grant agreement No 820445 and project name Quantum Internet Alliance. E.O acknowledges support from the Government of Spain (FIS2020-TRANQI and Severo Ochoa CEX2019-000910-S), Fundació Cellex, Fundació Mir-Puig, Generalitat de Catalunya (CERCA, AGAUR SGR 1381) and from the ERC AdGCERQUT.
\end{acknowledgments}

\medskip
P.C. and E.O. contributed equally to this work. 

\appendix
\section{Non-unit detection efficiency}
\label{sec:app_deteff}
We considered that the measurements are realized with non-photon-number-resolving detectors preceded by displacement operations in phase space. As explained in the main text, we assign the outcome $+1$ to a no-detection and $-1$ to a conclusive detection event. Given a state $\rho$ in a single bosonic mode with associated annihilation operator $a$ and creation operator $a^{\dagger}$, the probability $P_\mathrm{nc}$ to get an outcome $+1$ using a displacement with argument $\alpha$, is given by
\begin{equation}
P_\mathrm{nc}=\tr_a (D^{\dagger}(\alpha)\ketbra{0}{0}D(\alpha)\rho).
\end{equation}
In the case where the detector has a finite efficiency, we can model the detector inefficiency with a beam splitter having a transmission $\eta$, that is
\begin{equation}
P_\mathrm{nc}=\tr_{a} \Big(\ketbra{0}{0} \tr_{c}(U_\eta (D(\alpha)\rho D^{\dagger}(\alpha))\otimes \ketbra{0}{0}_{c} U_\eta^{\dagger})\Big)
\end{equation}
with $U_\eta = e^{\varphi(a^{\dagger}c-c^{\dagger}a)}$ for $\eta=\cos^2(\varphi)$, and the auxiliary mode described
by $c$ and $c^{\dagger}$, being initially empty. Using $U_\eta D(\alpha)= U_\eta D(\alpha)U^{\dag}_\eta U_\eta$ together with $U_\eta D(\alpha) U_\eta^{\dagger}=D_a(\alpha\sqrt{\eta})D_c(\alpha\sqrt{1-\eta})$, we end up with
\begin{equation}
\begin{split}
P_\mathrm{nc}&=\tr_{a}\Big(D^{\dagger}_a(\alpha\sqrt{\eta})\ketbra{0}{0}D_a(\alpha\sqrt{\eta}) \\
&\qquad\quad\cdot \tr_c(U_\eta \rho\otimes \ketbra{0}{0}_{c} U_\eta^{\dagger})\Big),
\end{split}
\end{equation}
where the displacement on mode $c$ has been traced out. This means that we can model the detection inefficiency as loss operating on the measured state if the amplitude of the displacement operation is changed accordingly. Hence, the fact that we consider detectors with unit efficiencies is still a valid description of our measurement apparatus, where we do not need any assumptions on our state nor on the efficiency of our detectors.

\section{Genuine multipartite entanglement witness}
\label{sec:app_witness}

Here we show how the calculation of the biseparable bound of our witness can be reduced to an single parameter optimization of the maximum eigenvalue of a $N\times N$-matrix. We start with the witness operator as presented in Eq.~\eqref{eq:witness}
\begin{equation}
\label{eq:witness_explicit}
    \hat{\mathcal{W}}= \hat{\mathcal{O}}_{\bm \alpha} + M_{n\leq 1}   -  N(N-1) \Pi_{n\geq 2} - \mu E_{n\geq 2},
\end{equation}
where $M_{n\leq 1}$ is an operator in the sector with not more than one photon and $E_{n\geq 2}\geq 0$ is in the sector with more than two photons. As argued in the main text, all our observables are block diagonal with respect to the total number of photons. In particular, $\hat{\mathcal{O}}_{\bm \alpha} =\hat{\mathcal{O}}_{\bm \alpha}^{n\leq 1} \oplus \hat{\mathcal{O}}_{\bm \alpha}^{n \geq 2}$. Furthermore, $\|\hat{\mathcal{O}}_{\bm \alpha}^{n \geq 2}\|\leq \|\hat{\mathcal O}\| = N(N-1)$ implies $\hat{\mathcal{O}}_{\bm \alpha}^{n\geq 2} - N(N-1) \Pi_{n\geq 2} \leq 0$ and
\begin{equation}
\begin{split}
    \hat{\mathcal{W}} &= (\hat{\mathcal{O}}_{\bm \alpha}^{n\leq 1} + M_{n\leq 1}) \\
    &\quad \oplus\,(\hat{\mathcal{O}}_{\bm \alpha}^{n\geq 2} -  N(N-1)\Pi_{n\geq 2} - \mu E_{n\geq 2})
    \\
    &\leq  (\hat{\mathcal{O}}_{\bm \alpha}^{n\leq 1} + M_{n\leq 1}) \oplus (-\mu\, E_{n\geq 2}).
\end{split}
\end{equation}
To find the biseparable bound consider biseparable states of the form
\begin{equation}
    \ket{\Psi} = \ket{\Psi_1}_{G_1} \ket{\Psi_2}_{G_2},
\end{equation}
defined over a bipartition given by two disjoint subsets $G_1 \cup G_2 =\{1,\dots,N\}$ of the $N$ modes. The value that the witness takes over these states reads
\begin{equation}\label{eq: bound bisep app}\begin{split}
\bra{\Psi}  \hat{\mathcal{W}} \ket{\Psi}
& \leq \bra{\Psi} \hat{\mathcal{O}}_{\bm \alpha}^{n\leq 1} + M_{n\leq 1} \ket{\Psi}\\
& \quad - \mu \bra{\Psi} E_{n\geq 2} \ket{\Psi}.
\end{split}
\end{equation}
As $\mu \bra{\Psi} E_{n\geq 2} \ket{\Psi}$ is positive, to maximize the right-hand side one can restrict the consideration to states $\ket{\Psi_{1(2)}}$ with one photon at most. Hence, without loss of generality we take
\begin{equation}
\ket{\Psi_k}_{G_k} = v_0^{(k)} \ket{0}_{G_k} + \sum_{i=1|j_i\in G_k}^{|G_k|} v_{i}^{(k)} a^\dag_{j_i}\ket{0}_{G_k}.
\end{equation}
To compute the value of the witness on these states we need the explicit form of the operators $\hat{\mathcal{O}}_{\bm \alpha}^{n\leq 1}$ and $M_{n\leq 1}$. To compute $\hat{\mathcal{O}}_{\bm \alpha}^{n\leq 1}$ we use the restriction of $\sigma_{\alpha}$ to the subspace with not more than one photon $\{\ket{0},\ket{1}\}$
\begin{equation}
    \sigma_{\alpha}^{n\leq 1} = \begin{pmatrix}
     f(\alpha) & g(\alpha)\\
     g(\alpha) & h(\alpha)
    \end{pmatrix},
\end{equation}
where we assumed a real $\alpha$ and denoted 
\begin{align}
f(\alpha)& =\big(2 e^{-\alpha^2}-1\big), \\
g(\alpha) &=2 \alpha e^{-\alpha^2}, \\
h(\alpha) &= \big(2 \alpha ^2 e^{-\alpha^2}-1\big),
\end{align}
see e.g. \cite{Vivoli2015} for a derivation. For the product observable restricted to the subspace of interest we obtain
\begin{widetext}
\begin{equation}
\begin{split}
\left(\sigma_{\alpha_i}^{(i)}\otimes  \sigma_{\alpha_j}^{(j)}\right)_{n\leq 1} &= f(\alpha_i) f(\alpha_j)\ketbra{00}{00}_{i,j}+g(\alpha_i)g(\alpha_j)\big(\ketbra{01}{10}_{i,j}+\ketbra{10}{01}_{i,j}\big) \\
 &\quad +f(\alpha_i)h(\alpha_j)\ketbra{01}{01}_{i,j}+h(\alpha_i)f(\alpha_j)\ketbra{10}{10}_{i,j},
\end{split}
\end{equation}
remark that here and in the following $\sigma_{\alpha_i}^{(i)}\otimes  \sigma_{\alpha_j}^{(j)}$ refers to the phase-averaged operator.
Adding the identity on the remaining modes gives 
\begin{equation}
\begin{split}
  &\Big(\sigma_{\alpha_i}^{(i)} \otimes \sigma_{\alpha_j}^{(j)}\otimes \id_{\lnot \{i,j\}}\Big)_{n\leq 1}=f(\alpha_i) f(\alpha_j) \bigg(\ketbra{\bar 0}{\bar 0} +\sum_{k\neq i,j}\ketbra{1_k}{1_k} \bigg)  \\ 
  &\quad +\left(g(\alpha_i)g(\alpha_j)(\ketbra{01}{10}_{i,j}+\ketbra{10}{01}_{i,j}) +f(\alpha_i)h(\alpha_j)\ketbra{01}{01}_{i,j}+h(\alpha_i)f(\alpha_j)\ketbra{10}{10}_{i,j}\right) \otimes \ket{\bar 0}\bra{\bar 0}_{\lnot \{i,j\}},
\end{split}
\end{equation}
with $\ket{\bar 0}$ denoting the vacuum state of all involved modes and $\ket{1_k}=a_k^\dag \ket{\bar 0}$ denoting the state with one photon in mode $k$ and vacuum elsewhere, and
\begin{equation}
\begin{split}
 \hat{\mathcal{O}}_{\bm \alpha}^{n\leq 1} &=\sum_{i\neq j}\left(\sigma_{\alpha_i}^{(i)}\otimes  \sigma_{\alpha_j}^{(j)}\otimes \id_{\lnot \{i,j\}}\right)_{n\leq 1}\\
 & = \sum_{i\neq j} f(\alpha_i) f(\alpha_j) \bigg(\ketbra{\bar 0}{\bar 0} +\sum_{k\neq i,j}\ketbra{1_k}{1_k} \bigg) + \sum_{i\neq j} g(\alpha_i)g(\alpha_j) \left( \ketbra{1_i}{1_j} +\ketbra{1_j}{1_i}\right)\\
 &\qquad +\sum_{i\neq j} \big(h(\alpha_i)f(\alpha_j)  \ketbra{1_i}{1_i} +h(\alpha_j)f(\alpha_i)  \ketbra{1_j}{1_j}\big).
\end{split}
\end{equation}
For $M_{n\leq 1}$  we chose
\begin{equation}
\label{eq:M1_explicit}
\begin{split}
 M_{n\leq 1} =\lambda\ket{\bar{0}}\bra{\bar{0}}-\sum_{i\neq j} f(\alpha_i) f(\alpha_j)\bigg(\ketbra{\bar 0}{\bar 0} + \sum_{k\neq i,j} \ketbra{1_k}{1_k}\bigg),
\end{split}
\end{equation}
with a positive real parameter $\lambda$ that one can tune, such that
\begin{equation}\begin{split}
   \hat{\mathcal{O}}_{\bm \alpha}^{n\leq 1}+   M_{n\leq 1} &= \lambda\ket{\bar{0}}\bra{\bar{0}} + \sum_{i\neq j} g(\alpha_i)g(\alpha_j) \left( \ketbra{1_i}{1_j} +\ketbra{1_j}{1_i}\right) \\
   &\quad +\sum_{i\neq j} \big(h(\alpha_i)f(\alpha_j)  \ketbra{1_i}{1_i} +h(\alpha_j)f(\alpha_i)  \ketbra{1_j}{1_j} \big) .
\end{split}
\end{equation} 
Finally for $E_{n\geq 2}$ we take the operators corresponding to the probability to find photons in more than one mode.
We rewrite $v_0^{(1)}=c_a$ and $v_0^{(2)}=c_b$, where we use the short notation $c_{a(b)}=\cos\big(a(b) \big)$ and $s_{a(b)}=\sin\big(a(b)\big)$, such that $\bm v^{(1)} = \binom{c_a}{s_a \bm v'}$ and $\bm v^{(2)} = \binom{c_b}{s_b \bm w'}$ with normalized $\bm v'$ and $\bm w'$ which leads to
\begin{equation}
    \ket{\Psi}=c_a c_b \ket{\bar{0}} + \sum_{i\in G_1} s_a c_b v^{\prime}_i \ket{1_i} + \sum_{j\in G_2} c_a s_b w^{\prime}_j \ket{1_j} +\sum_{i\in G_1}\sum_{j\in G_2}s_a s_b w^{\prime}_j v^{\prime}_i \ket{1_i}_{G_1}\ket{1_j}_{G_2},
\end{equation}
where $\ket{1_j}$ stands for one photon in mode $j$ and vacuum in all the other modes. We thus have by Eq.~\eqref{eq: bound bisep app}
\begin{equation}
\label{exp}
   \bra{\Psi}  \hat{\mathcal{W}} \ket{\Psi}\leq\lambda c_a^2 c_b^2-\mu s_a^2 s_b^2+2\big(\sum_{i\neq j}g(\alpha_i) g(\alpha_j)L_iL_j+f(\alpha_i)h(\alpha_j)L_j^2\big),
\end{equation}
where $\bm L= \binom{c_a s_b \bm w'}{s_a c_b \bm v'}$. We rewrite Eq. \eqref{exp} in a matrix form
\begin{equation}
\label{opti}
\begin{split}
\bra{\Psi}  \hat{\mathcal{W}} \ket{\Psi}&\leq  \lambda  c_a^2 c_b^2 - \mu s_a^2 s_b^2 + \bm L^T 
    \begin{pmatrix}
    M_w & M_c \\
    M_c^T & M_v
    \end{pmatrix} \bm L. 
\end{split}
\end{equation}
By arranging the entries of the matrix according to the defined bipartition, i.e. the block spanning the first $|G_2|$ rows and columns describes the modes in $G_2$, we can explicitly write down the matrix
\begin{equation}
\begin{split}
\begin{pmatrix}
  M_w & M_c \\
  M_c^T & M_v
 \end{pmatrix}
= 2 \begin{pmatrix}
  \sum_{i\neq 1}f(\alpha_i)h(\alpha_1) & g(\alpha_1)g(\alpha_2) & \cdots & g(\alpha_1)g(\alpha_N) \\
  g(\alpha_2)g(\alpha_1) & \sum_{i\neq 2}f(\alpha_i)h(\alpha_2) & \cdots & g(\alpha_2)g(\alpha_N) \\
  \vdots  & \vdots  & \ddots & \vdots  \\
  g(\alpha_N)g(\alpha_1) & g(\alpha_N)g(\alpha_2) & \cdots & \sum_{i\neq N}f(\alpha_i)h(\alpha_N) 
\end{pmatrix}.
\end{split}
\end{equation}
We finally can write
\begin{equation}
\begin{split}
    \bra{\Psi}  \hat{\mathcal{W}} \ket{\Psi} &\leq \lambda  c_a^2 c_b^2 -\mu s_a^2 s_b^2
    + \binom{c_a s_b \bm w^{\prime}}{s_a c_b \bm v^{\prime}}^T 
    \begin{pmatrix}
    M_w & M_c \\
    M_c^T & M_v
    \end{pmatrix} 
    \binom{c_a s_b \bm w^{\prime}}{s_a c_b \bm v^{\prime}} \\
    &= \lambda  c_a^2 c_b^2 - \mu s_a^2 s_b^2
    + \binom{s_b \bm w^{\prime}}{ c_b \bm v^{\prime}}^T 
    \begin{pmatrix}
    c_a^2 M_w &  c_a s_a M_c \\
     c_a s_a M_c^T &  s_a^2 M_v
    \end{pmatrix} 
    \binom{ s_b \bm w^{\prime}}{ c_b \bm v^{\prime}} \\
   & = \binom{s_b \bm w^{\prime}}{ c_b \bm v^{\prime}}^T
   \begin{pmatrix}
    c_a^2 M_w -s_a^2\mu \id &  c_a s_a M_c \\
     c_a s_a M_c^T &  s_a^2 M_v + c_a^2 \lambda \id
    \end{pmatrix} 
    \binom{ s_b \bm w^{\prime}}{ c_b \bm v^{\prime}}\\
     & \leq \text{max\,eig}\begin{pmatrix}
    c_a^2 M_w -s_a^2\mu \id  &  c_a s_a M_c \\
     c_a s_a M_c^T &  s_a^2 M_v + c_a^2 \lambda \id
    \end{pmatrix}
    =\text{max\,eig} \left( \mathds{M}(\lambda, \mu, \bm \alpha, a) \right),
\end{split}
\end{equation}
where in the last line "max eig" denotes the maximum eigenvalue of the Hermitian matrix $\mathds{M}(\lambda, \mu, \bm \alpha, a)$.

The biseparable bound is thus given by the optimization
\begin{equation}
    \widetilde{w}_\text{bisep} = \max_{G_1,G_2} w_{G_1,G_2} 
    = \max_{G_1,G_2} \max_{a\in [0,2\pi]} \text{max\,eig} \left( \mathds{M}(\lambda, \mu, \bm \alpha, a) \right)
\end{equation}
for well-chosen $\lambda$ and $\mu$. One notes that the optimization can be restricted to $a\in [0,\frac{\pi}{2}]$, as the transformation $(c_a^2,s_a^2, c_a s_a)\to (c_a^2,s_a^2, -c_a s_a)$ only changes the sign of the off-diagonal blocks of the matrix $\mathds{M}\to \begin{pmatrix}
 1& \\
  &-1
\end{pmatrix}\mathds{M}\begin{pmatrix}
 1& \\
  &-1
\end{pmatrix}$ which does not change its spectrum, as $ \begin{pmatrix}
 1& \\
  &-1
\end{pmatrix}$ 
is an orthogonal matrix (basis change).
\end{widetext}

\section{Measuring the witness }
\label{sec:app_measuring}
In this section, we explain how we estimate the expected value of the witness
\begin{equation}
\begin{split}
\hat{\mathcal{W}}_{\bm \alpha} = \hat{\mathcal{O}}_{\bm \alpha} + M_{n\leq 1}   -  N(N-1) \Pi_{n\geq 2} - \mu E_{n\geq 2}
\end{split}
\end{equation}
on the state $\rho$ prepared in the experiment.
It can be estimated from two observables, the one with displacement $\hat{\mathcal{O}}_{\bm \alpha}$ and one without 
$ M_{n\leq 1}   -  N(N-1) \Pi_{n\geq 2} - \mu E_{n\geq 2}$.

In our case, given the limited number of detectors we actually use three different observables, because we split $\Pi_{n\geq 2}$ in two parts. To do so we note that the probability that any $N$-mode state $\rho$ contains two or more photons satisfies
$P_{n\geq 2}( \rho )= \tr (\Pi_{n\geq 2}\,  \rho) \leq p_*$, where 
\begin{equation}\label{eq: pstar bound}
    p_* = \sum_{n=2}^N P_\text{click}^n+\sum_{i=1}^N p_{i}^*.
\end{equation}
Here, $P_\text{click}^n$ is the joint probability that $n$ detectors click and all the other detectors do not click when measured without displacement and $p_{i}^*$ is an upper bound on the probability of having two or more photons in mode $i$, that we associate to an observable $\Pi_{n\geq 2}^{(i)}$. The value for $p_{i}^*$ can be obtained in practice by measuring the probability of coincidence after a 50/50 beam splitter in mode $i$. On the level of the operators, we can write 
\begin{equation}
    \Pi_{n\geq 2} \leq E_{n\geq 2} + \sum_{i=1}^N \Pi_{n\geq 2}^{(i)}
\end{equation}
with two parts that we measure independently.
For the witness this implies 
\begin{equation}
\begin{split}
\hat{\mathcal{W}}&\geq \, \hat{\mathcal{O}}_{\bm \alpha} + M_{n\leq 1}   -  (N(N-1)+\mu) E_{n\geq 2} \\
&\quad - N(N-1)\sum_{i=1}^N \Pi_{n\geq 2}^{(i)}.
\end{split}
\end{equation}
 
It is already clear how the observables $ \hat{\mathcal{O}}_{\bm \alpha}$ and $\sum_{i=1}^N \Pi_{n\geq 2}^{(i)}$ can be measured, so let us now focus on the remaining terms. First, we note that
\begin{equation}\begin{split}
    - f(\alpha_i) f(\alpha_j)\bigg(\ketbra{\bar 0}{\bar 0} + \sum_{k\neq i,j} \ketbra{1_k}{1_k}\bigg) \\
    \geq -  \max\{f(\alpha_i) f(\alpha_j),0\} \ketbra{00}{00}_{ij}
\end{split}
\end{equation}
because the probability that there is no photon in the state or only one photon in some mode $k\neq i,j$, given by the POVM element $\ketbra{\bar 0}{\bar 0} + \sum_{k\neq i,j} \ketbra{1_k}{1_k}$, is lower than the probability that there is no photons in the modes $i$ and $j$  given by $\ketbra{00}{00}_{ij}$. We thus obtain
\begin{equation}
   \begin{split}
      M_{n\leq 1}  \geq \lambda \ketbra{\bar 0}{\bar 0}   - \sum_{i\neq j} \max\{f(\alpha_i) f(\alpha_j),0\} \ketbra{00}{00}_{ij}.
   \end{split} 
\end{equation}
In the experiment we do not have full information on the value of $\bm \alpha$ in a particular round, but rather a range of possible values ${\bm \alpha} =(\alpha_1,\dots, \alpha_N) \in A$. Therefore, the following bound will be useful
\begin{align}
\label{eq:M_app}
    M_{n\leq 1}  &\geq  \bar{M}_{n\leq 1},\\
    \bar{M}_{n\leq 1} &= \lambda \ketbra{\bar 0}{\bar 0}   - \sum_{i\neq j} F_{ij} \ketbra{00}{00}_{ij}, \\
\begin{split}
    F_{ij} &= \max_{{\bm \alpha} \in A}  {\max}\{f(\alpha_i) f(\alpha_j),0\} \\
    & = {\max} \{0,  \max_{{\bm \alpha} \in A} f(\alpha_i) f(\alpha_j) \}.
\end{split}
\end{align}
To summarize we have shown that
\begin{align}
\label{eq:W_app}
\hat{\mathcal{W}}_{\bm \alpha} \geq \overline{\mathcal{W}}_{\bm \alpha}  &= \hat{\mathcal{O}}_{\bm \alpha} + \mathcal{Z} - N(N-1)\Sigma_{n\geq 2}, \\ 
\begin{split}
\mathcal{Z} & = \lambda \ketbra{\bar 0}{\bar 0}  - \sum_{i\neq j} F_{ij} \ketbra{00}{00}_{ij}
\\
& \quad -(N(N-1)+\mu) E_{n\geq 2}, 
\end{split}
\\
\Sigma_{n\geq 2} & =  \sum_{i=1}^N \Pi_{n\geq 2}^{(i)}.
\end{align}

In the experiment the values of the observables $\hat{\mathcal{O}}$, $\mathcal{Z}$ and $\Sigma_{n\geq 2}$ are measured independently. The observable $\Sigma_{n\geq 2}$ is independent of the displacement amplitudes $\bm \alpha$ while $\hat {\mathcal{O}}_{\bm \alpha}$ is physically determined by $\bm \alpha$. The observable $\mathcal{Z}$ is computed using the knowledge of the range $A$ of possible values ${\bm \alpha} \in A$ for the coefficients $F_{ij}$, while the underlying physical measurement is performed without displacements. 

To analyze the experimental data, we assume that the state preparation is identical in each run of the experiment, such that is the same $N$-mode state $\rho$ is prepared repeatedly. On the other hand, the displacement amplitudes are subject to controlled fluctuations ${\bm \alpha} \in A$, with the possible range $A$ determined experimentally, see Appendix~\ref{sec:app_experiment}. Nevertheless, each round $k$ can be associated to some (unknown) value ${\bm \alpha}_k$. To prove that the state is GME it is sufficient to show that
\begin{equation}
\frac{1}{n}\sum_{k=1}^n \left \langle \hat{\mathcal{W}}_{{\bm \alpha}_k} -\widetilde{w}_\textrm{bisep}({\bm \alpha}_k) \right \rangle_\rho \!\!>0
\end{equation}
with $\langle X\rangle_\rho = \tr (X\rho)$, since each ${\bm \alpha}_k$ yields a valid GME-witness. By defining the worst-case biseparable bound
\begin{equation}
\label{eq:bisepmax}
    w_\textrm{bisep}^\textrm{max} = \max_{{\bm \alpha}\in A} \, \widetilde{w}_\textrm{bisep}({\bm \alpha}),
\end{equation}
and using the bound~\eqref{eq:W_app}, we can relax the GME condition to
\begin{equation}\label{eq: app average viol}
\frac{1}{n}\sum_{k=1}^n \left \langle \hat{\mathcal{O}}_{{\bm \alpha}_k} + \mathcal{Z} - N(N-1)\Sigma_{n\geq 2} - w_\textrm{bisep}^\textrm{max} \right \rangle_\rho \!\!>0,
\end{equation}
where the left-hand side is a lower bound on $\frac{1}{n}\sum_{k=1}^n \left \langle \hat{\mathcal{W}}_{{\bm \alpha}_k} -w_\textrm{bisep}({\bm \alpha}_k) \right \rangle_\rho$.

Before analyzing the statistical significance of our data, let us briefly sketch how GME can be guaranteed in the asymptotic limit $n\to \infty$. Then the average values of the observables $\mathcal Z$ and $\Sigma_{n\geq 2}$ converge to their expected values $\langle \mathcal{Z} \rangle$ and $\langle\Sigma_{n\geq 2}\rangle$. Similarly, for the random variables $o_k$ as the value of $\hat{\mathcal{O}}_{{\bm \alpha}_k}$ observed in the round $k$ (where it is measured), the observed average converges to the average expected value 
\begin{equation}
\frac{1}{n}\sum_{k=1}^n o^{(k)} \to \frac{1}{n}\sum_{k=1}^n \mathds{E}(o^{(k)})  = \frac{1}{n}\sum_{k=1}^n \langle \hat{\mathcal{O}}_{{\bm \alpha}_k} \rangle_\rho,
\end{equation} 
by Hoeffding's  theorem (1963), see Appendix~\ref{sec:app_statistics}. Hence, Eq.~$\eqref{eq: app average viol}$ can be directly guaranteed from the data. Note that in practice to estimate $\Sigma_{n\geq 2}$ we do not measure $\Pi_{n\geq 2}^{(i)}$ for each mode. Instead, we assume that it has the same expected value for each mode, such that
\begin{equation}
\langle \Sigma_{n\geq 2} \rangle_\rho = N \langle \Pi_{n\geq 2}^{(1)}\rangle_\rho, 
\end{equation}
and only estimate $\langle \Pi_{n\geq 2}^{(1)}\rangle_\rho$.

\begin{table*}
\capstart
\centering
\small
\setlength\tabcolsep{6pt}
\begin{tabular}{c c c c c c c c c c c} 
\toprule
$N$ & $\lambda$ & $\mu$ & $\overline o$ & $\overline z$ & $\overline s$ & $n$ & $m$ & $\ell$ & $w_\mathrm{bisep}^\mathrm{max}$ & $p$-value \\[0.5ex]
\cmidrule{1-11}
4 & 2.73 & 102 & 1.1525 & 1.8417 & -0.0014 & 26747089 & 26755161 & 135905902 & 2.785 & \num{E-1952} \\
8 & 8.29 & 151 & 2.5762 & 5.9915 & -0.0024 & 27611104 & 27576602 & 365370348 & 8.358 & \num{E-87} \\
\bottomrule
\end{tabular}
\caption{\label{tab:Statistics} Evaluation of the witness for $N$ parties with parameters $\lambda$ and $\mu$ according to Eqs.~\eqref{eq:witness_explicit} and \eqref{eq:M1_explicit}. The mean values $\overline o$, $\overline z$ and $\overline s$ are associated to the observables $\hat{\mathcal{O}}_{\bm \alpha}$, $\mathcal{Z}$ and $-N^2(N-1)\Pi_{n\geq 2}^{(1)}$, respectively. The numbers $n$, $m$ and $\ell$ indicate the number of evaluations of $o$, $z$ and $s$. Together with the biseparable bound $w_\mathrm{bisep}^\mathrm{max}$ according to Eq.~\eqref{eq:bisepmax}, the $p$-value is calculated using Eq.~\eqref{eq:pvalue}. Note that in order to obtain a $p$-value of less than \num{E-10}, in the case of $N$=4 it would suffice to evaluate the observables $n=m=\ell=\num{4.9E5}$ times, corresponding to a total integration time of less than \SI{130}{s}, and for $N$=8 it would require $n=m=\ell=\num{7.8E6}$ evaluations which could be achieved in less than \SI{2100}{s}.}
\end{table*}
\section{Finite statistics analysis}
\label{sec:app_statistics}

In the experiment, three different measurements are performed, to each of which we associate a random variable. Let $o_k$ be the random variable given the value of $\hat{\mathcal{O}}_{{\bm \alpha}_k}$ observed in the round $k=1,\dots,n$ (when it is measured). Analogously, define $z_k$ associated to $\mathcal{Z}$ (for $k=1,\dots,m$) and $s_k$ associated to $-N^2(N-1)\Pi_{n\geq 2}^{(1)}$ (for $k=1,\dots,\ell$). Note that all the variable are independent, furthermore the variables $z_k$ and $s_k$ are also identically distributed (for each $k$). From the definition of the corresponding observables (their spectrum) one directly sees that
\begin{align}
      o_k &\in [-N(N-1), N(N-1)],\\
      z_k & \in [-\sum_{i\neq j}F_{ij}-N(N-1)-\mu,\lambda],\\
      s_k & \in [-N^2(N-1),0],
\end{align}
from which we define
\begin{align}
    \Delta_o &= 2N(N-1),\\
    \Delta_z &= \lambda + \sum_{i\neq j}F_{ij}+N(N-1)+\mu,\\
    \Delta_s &= N^2(N-1).
\end{align}
For each type of observables we define the average as
\begin{equation}
    \bar o = \frac{1}{n} \sum_{k=1}^n o_k.
\end{equation}

To analyze the statistical significance of our data, we use the following theorem by Hoeffding (1963) \cite{Hoeffding1963}: For any collection of independent random variables $x^{(1)},\dots, x^{(n)}$ with $x^{(k)}\in [a_k, a_k+\Delta_k]$ the following bound holds 
\begin{equation}\begin{split}
    &\textrm{P}\Big( \overline x-t \geq \mathds{E}\left(\overline x\right) \Big)\leq  \exp\left(-\frac{2 n^2 t^2}{\sum_{k=1}^n\Delta_k^2}\right)\\
&\textrm{for}\quad \overline x =\frac{1}{n} \sum_{i=1}^n x^{(k)}. 
\end{split}
\end{equation}

To apply to our data consider the situation where the observable $o_k\in [a,a+\Delta_o]$ are measured in $n$ rounds, $z_k\in [b,b+\Delta_z]$ are measured in $m$ rounds, and $s_k\in [c,c+\Delta_s]$ are measured in $\ell$ rounds, the above theorem implies  
\begin{equation}\label{eq: P bound}
    \textrm{P}\Big( \overline o +\overline z +\overline s -t \geq \mathds{E}\left(\overline o + \overline z +\overline s\right) \Big)\leq  e^{-\frac{2 (n+m+\ell)^2 t^2}{n\Delta_o^2+ m \Delta_z^2+\ell \Delta_s^2}}.
\end{equation}
Now consider any state $\rho_\textrm{bisep}$ that is not GME. We have shown that such a state does not violate the relaxed witness of Eq.~\eqref{eq: app average viol}. Thus, it gives rise to a collection of random variables, described in the beginning of the section, with
\begin{equation}\begin{split}
   0 &\geq
    \Big \langle  \frac{1}{n} \sum_{k=1}^n\hat{\mathcal{O}}_{{\bm \alpha}_k}  + \mathcal{Z} -  N(N-1) \Sigma_{n\geq 2} - w_\textrm{bisep}^\textrm{max}\Big \rangle_{\rho_\textrm{bisep}}\\
    & =\mathds{E}\left(\overline o + \overline z +\overline s -w_\textrm{bisep}^\textrm{max}\right)
\end{split}
\end{equation}
Then, by Eq.~\eqref{eq: P bound} the probability that the observed averages satisfy
\begin{equation}
\overline o +\overline z +\overline s - w_\textrm{bisep}^\textrm{max}  \geq t,
\end{equation}
i.e. a fake violation exceeding $t$ is observed due to statistical fluctuation,  is upper bounded by
\begin{equation}
\exp\left(-\frac{2 (n+m+\ell)^2 t^2}{n\Delta_o^2+ m \Delta_z^2+\ell \Delta_s^2}\right).
\end{equation}
Hence,  
\begin{equation}
\label{eq:pvalue}
p=\exp\left(-\frac{2 (n+m+\ell)^2 (\overline o +\overline z +\overline s - w_\textrm{bisep}^\textrm{max})^2}{n\Delta_o^2+ m \Delta_z^2+\ell \Delta_s^2}\right)
\end{equation}
can be interpreted as the $p$-value associated to our GME test. That is, $p$ is an upper bound on the probability that a state $\rho_\textrm{bisep}$ which is not GME produces a fake violation of $\overline o +\overline z +\overline s - w_\textrm{bisep}^\textrm{max}$ or higher. 

For the corresponding values in the experiment, see Tab.~\ref{tab:Statistics}.

\section{Experimental methods and characterization}
\label{sec:app_experiment}
To ensure high-purity heralded signal photons, we spectrally filter the heralding idler photons emitted from the PPKTP crystal by using a dense wavelength division multiplexer (DWDM) with a \SI{100}{GHz} passband at $\lambda_i=\SI{1546.12}{nm}$ (ITU channel 39). 
\begin{figure}
\capstart
\includegraphics[width = 0.96\columnwidth]{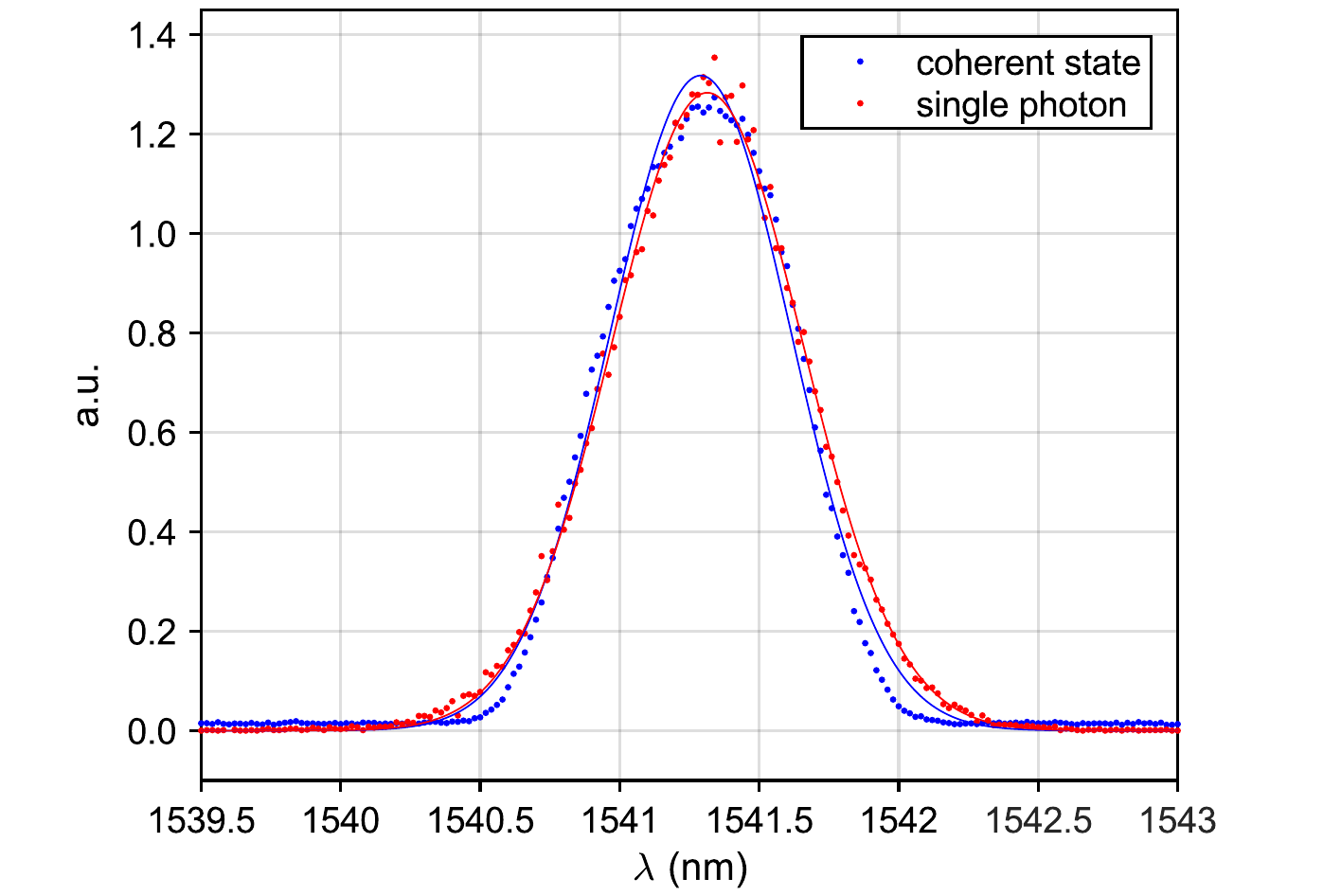}
\caption{\label{fig:app_overlap} Normalized measured spectra and Gaussian fit of single-photon and coherent states. The spectra are measured with a tunable grating filter with a FWHM of \SI{0.2}{nm} inserted before the SNSPD.}
\end{figure}

\begin{figure}
\capstart
\includegraphics[width = \columnwidth]{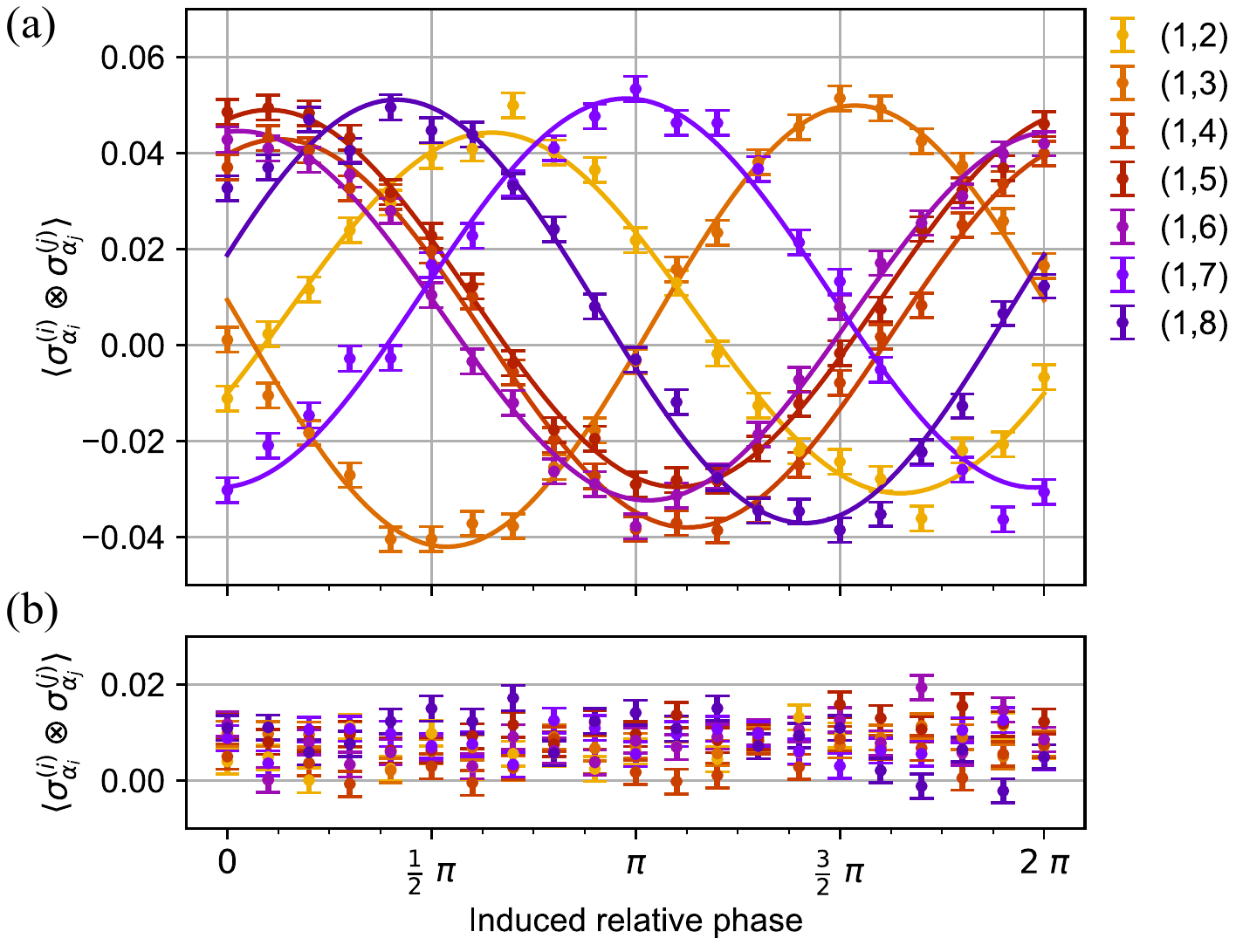}
\caption{\label{fig:phasescan} Phase alignment between mode 1 (reference) and all seven other modes. (a) The single-photon state is displaced with $\alpha \approx 0.83$ and the first segment of the electronic polarization controller is swept over its full range from $0$ to $2\pi$ in each non-reference mode. Sinusoidal curves are fitted to the data. (b) Same measurement without the presence of the single-photon state.}
\end{figure}

Besides high single-photon purity, the single-photon and coherent states need to have a good spectral and temporal overlap in order to perform the targeted displacement operation. The measurement of the spectral overlap is shown in Fig.~\ref{fig:app_overlap}.
The expected Hong-Ou-Mandel visibility due to the finite overlap of the fitted Gaussians is \SI{99.2}{\%} \cite{Mosley2007}.

\begin{figure}
\capstart
\includegraphics[width = \columnwidth]{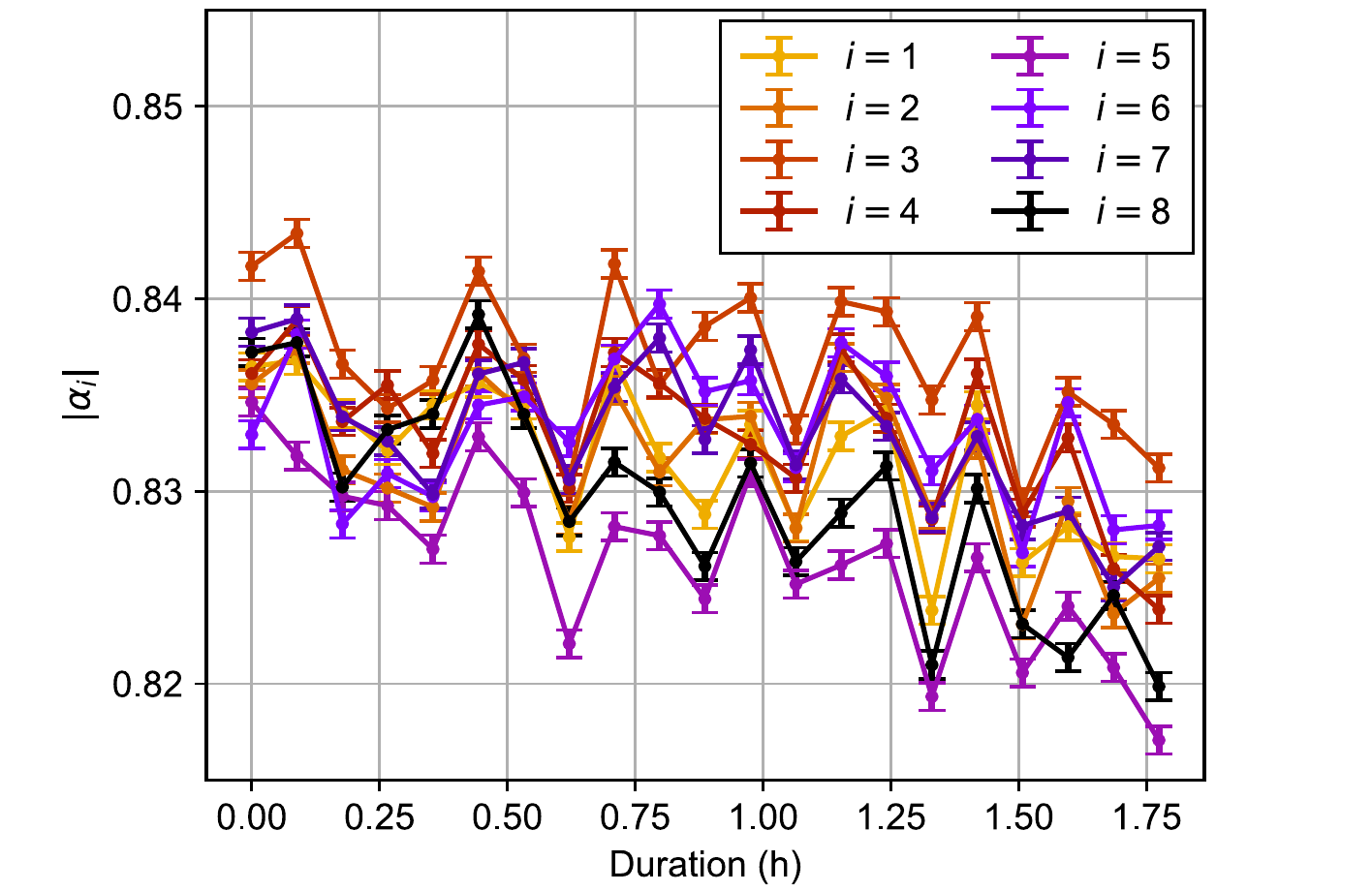}
\hspace{0.7cm}
\caption{\label{fig:app_alpha} Measurement of the displacement amplitudes $|\alpha_i|$ for each spatial output mode $i$ during the data acquisition. For each point, counts are acquired for \SI{1}{min}.}
\end{figure}

The temporal alignment between the single-photon and coherent states is done in the following way. The seed laser is switched to continuous mode and, together with the pulsed 
pump laser, difference frequency generation in both nonlinear crystals, PPKTP and PPLN, is used to generate coherent states at $\lambda_s$. First-order interference is observed at one of the output modes and the interference visibility is maximized by adjusting the delay of the motorized delay line.

\begin{figure}
\capstart
\includegraphics[width = \columnwidth]{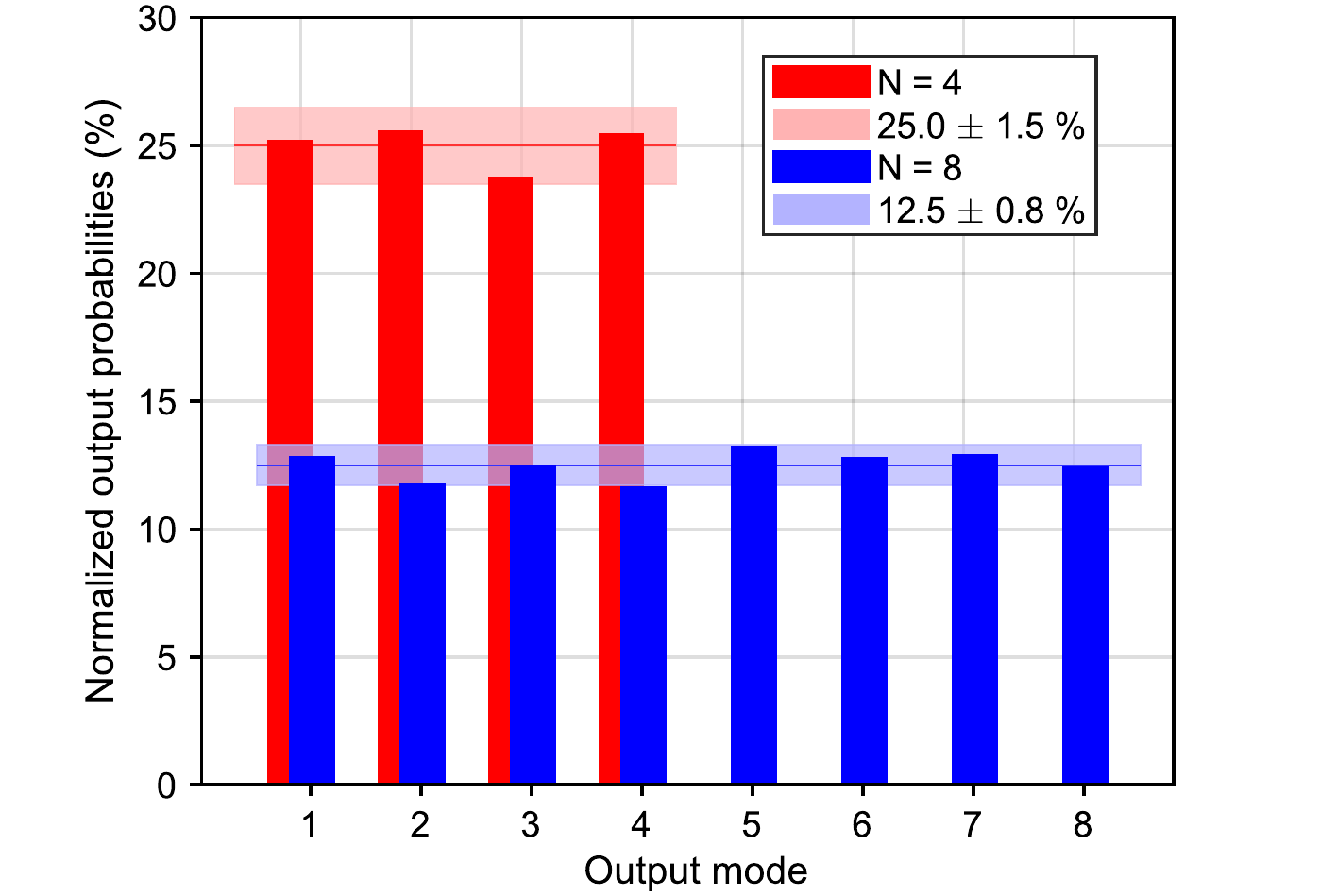}
\caption{\label{fig:app_balance} Normalized probabilities of detecting a photon in output mode $i\in\{1,...,N\}$ in the measurement without displacement.}
\end{figure}

In order to set the relative phases between single-photon and coherent states in each spatial mode, polarization is aligned with a polarization controller such that it enters the first segment of the electronic polarization controller on-axis which therefore allows for relative phase control locally, as shown in Fig.~\ref{fig:phasescan}. The phase is then set in each mode such that its relative phase to the reference mode is zero.

After the phase alignment, data is acquired in intervals, as described in the main text. 
The displacement amplitudes are obtained by measuring the coherent state in the absence of the single-photon state and assuming Poissonian count statistics (see Fig.~\ref{fig:app_alpha}). For the calculation of the expectation value of the witness (see Eqs.~\eqref{eq:M_app} and \eqref{eq:W_app}) and the biseparable bound according to Eq.~\eqref{eq:bisepmax}, all the values $\bm \alpha \in A = \{(\alpha_1,\dots, \alpha_8)\in \mathbb{R}^8 | \alpha_{i}^\mathrm{min} \leq \alpha_{i} \leq \alpha_{i}^\mathrm{max}\}$ are considered. Here, $\alpha_{i}^\mathrm{min}$ ($\alpha_{i}^\mathrm{max}$) denotes the minimum (maximum) value of $\alpha_i$ on mode $i$ during the measurement shown in Fig.~\ref{fig:app_alpha}.

To determine the balance of the generated state, the counts in each mode in the measurement without displacement are normalized, as shown in Fig.~\ref{fig:app_balance}. For the measurement where $N=4$, the mode balance is $25.0\pm 1.5\%$, where for $N=8$ a balance of $12.5\pm 0.8\%$ is achieved. 

\begin{table}
\capstart
\centering
\small
\setlength\tabcolsep{3pt}
\begin{tabular}{c c c c c} 
\toprule
$N$ & $P_\text{click}^1$ & $P_\text{click}^2$ & $P_\text{click}^3$ & $p_{1}^*$ \\[0.5ex]
\cmidrule{1-5}
4 & 0.3106(1) & $3.21(4)\cdot 10^{-4}$ & $1.5(8)\cdot 10^{-7}$ & $2.89(5)\cdot 10^{-5}$ \\
8 & 0.2678(1) & $2.96(4)\cdot 10^{-4}$ & $1.5(8)\cdot 10^{-7}$ & $5.4(1)\cdot 10^{-6}$ \\
\bottomrule
\end{tabular}
\caption{\label{tab:StateProbab} Measured probabilities $P_\text{click}^n$ that $n$ detectors click in the case of preparing an $N$-partite state. Further, $p_{1}^*$ is an upper bound on the probability of having more than one photon locally in mode 1.}
\end{table}

The values for the probabilities $P_\text{click}^n$ that $n$ detectors click when measuring the state are given in Tab.~\ref{tab:StateProbab}. Note that the probability $p_0$ indicated in Tab.~\ref{tab:ResultsWitness} are $p_0=P_\text{click}^0=1-\sum_{n\geq 1} P_\text{click}^n$.

\section{The effect of dark counts on the witness}
\label{sec:app_dark_counts_GME}

Here we explain how the presence of detector dark counts can be in included in our GME witness. 
We will show that by adding a constant term $2 N^2 (N-1) p_{dc}$ to the biseparable bound, the violation of this shifted witness (exactly as described in the main text, but with detectors subject to dark counts) allows one to conclude that the measured state is GME. Before we start, recall that a usual model of dark counts for non-photon-number-resolving (NPNR) detectors is a classical noise which changes the outcome "no-click" to "click" with probability $p_{dc}$. Thus, "turning on" the dark counts on a detector modifies the click/no-click probabilities to
\begin{equation}
(p_0^{dc}, p_c^{dc}) = ((1-p_{dc})p_0, p_{dc}\, p_0 + p_c).
\end{equation}

The starting point is to consider an experiment where the expected value of the witness $\ev{ \hat{\mathcal{W}}_{dc} }$ is estimated as described in the main text, but with detectors subject to dark counts. We now introduce a simple physical model that reproduces (almost) all the statistics of this experiment, but involves detectors without dark counts. To this end, consider a single-mode quantum channel $S_{dc}$ which does nothing with probability $1-p_{dc}$ and replaces a single mode state $\varrho$ with a very bright Fock state $\ket{M}$ with probability $p_{dc}$
\begin{equation}
    S_{dc}: \varrho \mapsto S_{dc}[\varrho]=(1-p_{dc}) \varrho + p_{dc} \ketbra{M}{M}.
\end{equation}
The state $S_{dc}[\varrho]$ is then measured with the measurement described in the main text. We now have to distinguish between different measurements that we treat separately. (1) The measurements of the witness and (2) the estimation of $p_*$. More precisely, we need to distinguish measurements with one detector per mode and g$^{(2)}$ measurements with two detectors per mode. 

In the case (1) all measurements involve a single lossy NPNR detector per mode, sometimes preceded by a displacement $D(\alpha)$. The probability distribution of the outcomes of this measurement is a mixture of two possibilities. With probability $1-p_{dc}$ the state was unchanged and the measurement is performed on the original state $\rho$ leading to the click/no-click probabilities $(p_0,p_c)$. With probability $p_{dc}$ the measurement is performed on the state $\ket{M}$, where we can always choose $M$ large enough such that in this case a "click" outcome is observed with certainty $p_c=1$. The overall outcome probabilities are thus given by 
\begin{equation}\begin{split}
(p_0^{dc}, p_c^{dc}) &= (1-p_{dc})(p_0, p_c) + p_{dc} (0,1)\\
   &=((1-p_{dc}) p_0, (1-p_{dc}) p_c + p_{dc})\\
   &= ((1-p_{dc})p_0, p_{dc} p_0 + p_c).
\end{split}
\end{equation}
Hence, the $N$-mode state $\tilde \rho=S_{dc}^{\otimes N}[\rho]$ measured with detectors without dark-counts reproduces the statistics of the $N$-mode state $\rho$ observed with detectors subject to dark counts whenever only one detector is used per mode. 

Let us now consider the estimation of $p_*$, which is an upper bound on the probability of having two or more photons in the state. As defined in Eq.~\eqref{eq: pstar bound}, $p_*$ is composed of two contributions. The first term $\sum_{n=2}^N P_\text{click}^n$ is the probability to observe clicks on more than two modes gathered with a single detector per mode. Hence, for this term the above argumentation holds $P_\text{click+dc}^n[\rho] = P_\text{click}^n[\tilde \rho]$. The other term $p_i^*$ is an upper bound on the probability of having two or more photons in a single mode. This is measured in a $g^{(2)}$ experiment -- a single mode state $\varrho$ is split on a 50/50 beam splitter, and each output is sent to a NPNR detector. The probability that the two detectors click $p_{cc}$ is bounded by the probability that $\varrho$ contains two or more photons $ p_{cc} \leq p_i^* \leq  2 p_{cc}$. And it is precisely the estimated $p_{cc}$, which is used to bound $p_*$. Let us now analyze how this probability is affected by dark counts. For a state $\varrho$ we have
\begin{equation}
\begin{split}
    p_{cc}^{dc} &= p_{cc} + p_{dc} (p_{0c}+p_{c0}) + p_{dc}^2 p_{00}\\
    &= p_{cc} + p_{dc} (p_{0c}+p_{c0}+ p_{00}) - p_{dc} p_{00}+ p_{dc}^2 p_{00}\\
    &= p_{cc} + p_{dc} (1-p_{cc}) - p_{00} p_{dc}(1-p_{dc}).
\end{split}
\end{equation}
Now let us analyze the effect of the channel $S_{dc}$ on this probability. For a state $\tilde \varrho = S_{dc}[\varrho]$ one has
\begin{equation}
\begin{split}
    \tilde p_{cc} &= (1-p_{dc})p_{cc} + p_{dc}\\
    & = p_{cc} + p_{dc}(1-p_{cc}) \\
    &= p^{dc}_{cc} + p_{00} p_{dc}(1-p_{dc})\\
    &\leq p^{dc}_{cc} + p_{dc}.
\end{split}
\end{equation}
Hence, for the state $\tilde \rho$ we get an upper bound 
\begin{equation}
    \tilde{p}_i^* \leq p_i^* + 2 p_{dc},
\end{equation}
where $p_i^*$ is the quantity estimated in the experiment with dark counts. Combining the above arguments and using Eq.~\eqref{eq: pstar bound} it follows that 
\begin{equation}\label{eq: bound p tilde star}
\tilde p_* \leq p_*^{dc} + 2 N p_{dc}
\end{equation}
with $p_*^{dc}$ the $p_*$ estimated in the experiment with dark counts, is a valid upper bound on two (and more) photon contributions in the state $\tilde \rho = S_{cd}^{\otimes N}[\rho]$.\\

To summarize, the value of the witness estimated on an $N$-mode state $\rho$ with detectors subject to dark counts
\begin{equation}
\label{eq:witness_dc}
    \ev{ \hat{\mathcal{W}}_{dc} } = \tr(\hat{\mathcal{W}}_{dc} \, \rho ) = \tr(\hat{\mathcal{W}} \, S_{dc}^{\otimes N}[\rho])
\end{equation}
corresponds to the values of the original witness (without dark counts) estimated on the state $S_{dc}^{\otimes N}[\rho]$. On the other hand, the biseparable bound 
for the state $ S_{dc}^{\otimes N}[\rho]$ satisfies
\begin{equation}\begin{split}
\tilde w_\mathrm{bisep}^\mathrm{max} &= w_\text{bisep} + \tilde p_* N(N-1)\\
&\leq w_\text{bisep} + p_*^{dc} N(N-1) + 2 N^2(N-1) p_{dc},
\end{split}
\end{equation}
where we used Eq.~\eqref{eq: bound p tilde star}. Here, $w_\mathrm{bisep}^\mathrm{max+dc} = w_\text{bisep} + p_*^{dc} N(N-1)$ is the biseparable bound estimated in the real experiment (state $\rho$ and dark counts). We can thus conclude that observing
\begin{equation}
\ev{ \hat{\mathcal{W}}_{dc} }- w_\mathrm{bisep}^\mathrm{max+dc}  - 2 N^2 (N-1) p_{dc} \geq 0 
\end{equation}
implies that the state $ S_{dc}^{\otimes N}[\rho]$ is  GME by our main result. Since the channel $S_{dc}^{\otimes N}$ describes noise acting locally on each mode and cannot create entanglement, the $N$-mode state $\rho$ is also GME. This concludes our argument showing that if the detectors used in the experiment suffer from dark counts (that we did not include in their mode), the procedure described in the main text still allows to prove the GME of the measured state, but the observed violation $\ev{ \hat{\mathcal{W}}_{dc} }- w_\mathrm{bisep}^\mathrm{max+dc}$ has to exceed $2 N^2 (N-1) p_{dc}$. Note that since $\ev{ \hat{\mathcal{W}}_{dc} }- w_\mathrm{bisep}^\mathrm{max+dc}$ scales as $N^2$ in general, the penalty terms accounting for dark-count scales as $2 N p_{dc}$. \\

In the experiment with $N=8$, we measured $p_{dc}=\num{1.16(29)E-6}$ for the detector with the highest dark count rate. Therefore, the measured witness violation of $\ev{ \hat{\mathcal{W}}_{dc} }- w_\mathrm{bisep}^\mathrm{max+dc}=0.207(4)$ is reduced by $2 N^2 (N-1) p_{dc} = 0.0010(3)$, which still certifies GME with a $p$-value of \num{E-87}. \\

For the analysis of the scalability of the presented witness including detector dark counts, we consider a heralded single photon generated by a SPDC source, which state after heralding can be well approximated by
\begin{equation}
\rho=\frac{1}{1+p}(\ketbra{1}{1}+p\ketbra{2}{2})
\end{equation}
Since losses on the state commute with the beam splitter one can directly apply losses on $\rho$ to account for finite efficiency and obtain $\rho_{\eta}$, which transforms to $\rho_{\eta}^{BS}$ after the $N$-mode beam splitters. The idea in the following is to only consider the reduced density matrix to two modes
\begin{equation}
\sigma_{\{1,2\}}= \tr_{\{3,\dots ,N\}}(\rho_{\eta}^{BS}),
\end{equation}
which we use to compute
\begin{equation}
\begin{split}
    &S_{dc}^{\otimes 2}[\sigma_{\{1,2\}}] = (1-p_{dc})^2\sigma_{\{1,2\}} \\
    &\quad+ p_{dc}(1-p_{dc})(\sigma_{1}\otimes \ketbra{M}{M}+\ketbra{M}{M}\otimes\sigma_{1} )\\
    &\quad+ p_{dc}^2\ketbra{M,M}{M,M}.
\end{split}
\end{equation}
Since we consider a perfectly balanced beam splitter, we can evaluate almost all the terms in Eq.~\eqref{eq:witness_dc} only using $S_{dc}^{\otimes 2}[\sigma_{1,2}]$, where the only assumptions we make is that the 3-click events are always negligible over the 2-click events, which holds true for our state.

\begin{figure}
\capstart
\includegraphics[width = 0.92\columnwidth]{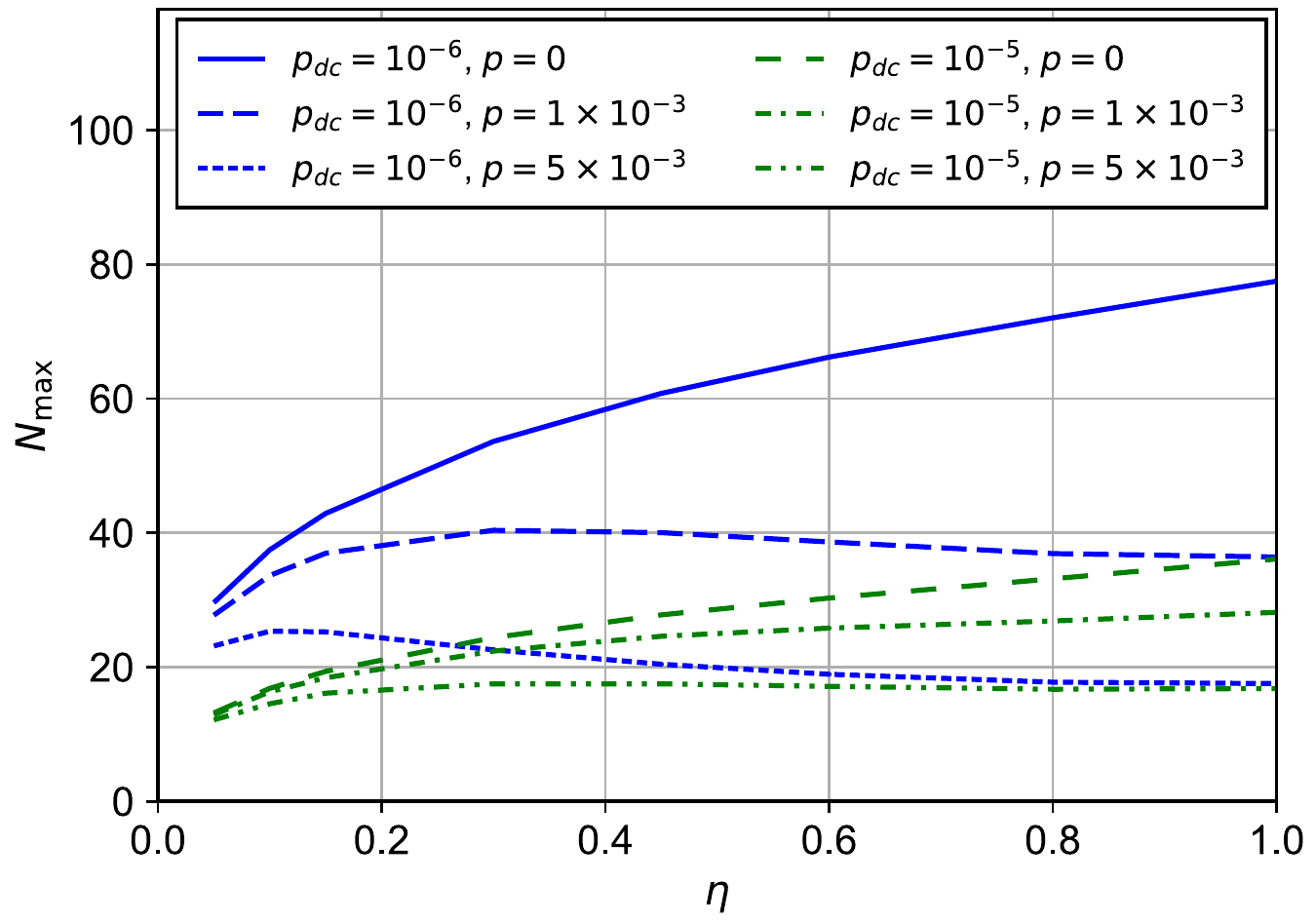}
\caption{\label{fig:app_scalability} Maximum number of parties $N_\text{max}$ for which the witness is still violated as a function of $\eta$ for a state $\rho_\eta$. The scaling behavior is shown for input states $\rho=(\ketbra{1}{1}+p\ketbra{2}{2})/(1+p)$ with $p\in\{0,\num{1E-3},\num{5E-3}\}$ that are measured with detectors suffering from dark counts with probabilities per heralding event of $p_{dc}\in\{\num{E-6},\num{E-5}\}$.}
\end{figure}
In Fig.~\ref{fig:app_scalability}, we show the maximum number of parties $N_\text{max}$, for which the value $\ev{ \hat{\mathcal{W}}_{dc} } - w_\mathrm{bisep}^\mathrm{max+dc}- 2 N^2 (N-1) p_{dc}$ is still positive, as a function of $\eta$ for a state $\rho_\eta^{BS}$. In the calculation, the displacement amplitude for the measurement is set to $\alpha=\sqrt{\ln{2}}\approx 0.83$ for each party, which is experimentally the most robust to fluctuations in $\alpha$. In the experiment, the probability of heralding a two-photon Fock state is $p\approx \num{4.9E-3}$ and the detector with the highest dark count rate has a dark count probability per heralding event of $p_{dc}=\num{1.16(29)E-6}$, which is approximated in the plot by the line $(p,p_{dc}) = (\num{5E-3},\num{E-6})$. We notice that in this case for $\eta > 0.05$, we are able to detect GME for more than $N=17$ parties. We further note that for higher dark count probabilities $N_\text{max}$ decreases, and in the case of $p=0$ the influence of dark counts can be substantial. The witness could be improved by using a different detector model including dark counts, which we leave for further work.

\printbibliography

@article{Bruno2014,
author={Bruno, Natalia and Martin, Anthony and Thew, Robert T.},
journal = {Optics Communications},
pages = {17--21},
publisher = {Elsevier},
title = {{Generation of tunable wavelength coherent states and heralded single photons for quantum optics applications}},
doi = {10.1016/j.optcom.2014.02.025},
volume = {327},
year = {2014}
}

@article{Caloz2018,
author = {Caloz, Misael and Perrenoud, Matthieu and Autebert, Claire and Korzh, Boris and Weiss, Markus and Sch{\"{o}}nenberger, Christian and Warburton, Richard J. and Zbinden, Hugo and Bussi{\`{e}}res, F{\'{e}}lix},
journal = {Applied Physics Letters},
number = {6},
title = {{High-detection efficiency and low-timing jitter with amorphous superconducting nanowire single-photon detectors}},
doi = {10.1063/1.5010102},
volume = {112},
year = {2018}
}

@article{Caspar2020,
  title = {Heralded Distribution of Single-Photon Path Entanglement},
  author = {Caspar, Patrik and Verbanis, Ephanielle and Oudot, Enky and Maring, Nicolas and Samara, Farid and Caloz, Misael and Perrenoud, Matthieu and Sekatski, Pavel and Martin, Anthony and Sangouard, Nicolas and Zbinden, Hugo and Thew, Robert T.},
  journal = {Phys. Rev. Lett.},
  volume = {125},
  issue = {11},
  pages = {110506},
  numpages = {6},
  year = {2020},
  publisher = {American Physical Society},
  doi = {10.1103/PhysRevLett.125.110506}
}

@article{Chou2005,
author = {Chou, Chin-Wen and {de Riedmatten}, Hugues and Felinto, Daniel and Polyakov, Sergey V. and Van Enk, Steven J. and Kimble, H. Jeff},
journal = {Nature},
number = {7069},
pages = {828--832},
pmid = {16341008},
title = {{Measurement-induced entanglement for excitation stored in remote atomic ensembles}},
doi = {10.1038/nature04353},
volume = {438},
year = {2005}
}

@article{Delteil2016,
author = {Delteil, Aymeric and Sun, Zhe and Gao, Wei Bo and Togan, Emre and Faelt, Stefan and Imamoglu, Ata{\c{c}}},
journal = {Nature Physics},
number = {3},
pages = {218--223},
title = {{Generation of heralded entanglement between distant hole spins}},
doi = {10.1038/nphys3605},
volume = {12},
year = {2016}
}

@article{Duan2001,
author = {Duan, L.-M. and Lukin, Mikhail D. and Cirac, J. Ignacio and Zoller, Peter},
journal = {Nature},
number = {6862},
pages = {413--418},
title = {{Long-distance quantum communication with atomic ensembles and linear optics}},
doi = {10.1038/35106500},
volume = {414},
year = {2001}
}

@article{Duer2000,
author = {D{\"u}r, Wolfgang and Vidal, Guifre and Cirac, J. Ignacio},
journal = {Physical Review A},
number = {6},
pages = {062314},
title = {{Three qubits can be entangled in two inequivalent ways}},
doi = {10.1103/PhysRevA.62.062314},
volume = {62},
year = {2000}
}

@article{Gottesman2012,
author = {Gottesman, Daniel and Jennewein, Thomas and Croke, Sarah},
journal = {Physical Review Letters},
number = {7},
pages = {070503},
title = {{Longer-Baseline Telescopes Using Quantum Repeaters}},
doi = {10.1103/PhysRevLett.109.070503},
volume = {109},
year = {2012}
}

@article{Graefe2014,
author = {Gr{\"{a}}fe, Markus and Heilmann, Ren{\'{e}} and Perez-Leija, Armando and Keil, Robert and Dreisow, Felix and Heinrich, Matthias and Moya-Cessa, Hector and Nolte, Stefan and Christodoulides, Demetrios N. and Szameit, Alexander},
issn = {17494893},
journal = {Nature Photonics},
number = {10},
pages = {791--795},
publisher = {Nature Publishing Group},
title = {{On-chip generation of high-order single-photon W-states}},
doi = {10.1038/nphoton.2014.204},
volume = {8},
year = {2014}
}

@article{Hoeffding1963,
author = {Hoeffding, Wassily},
journal = {Journal of the American Statistical Association},
number = {301},
pages = {13--30},
title = {{Probability Inequalities for Sums of Bounded Random Variables}},
doi = {10.1080/01621459.1963.10500830},
volume = {58},
year = {1963}
}

@article{Humphreys2018,
author = {Humphreys, Peter C. and Kalb, Norbert and Morits, Jaco P. J. and Schouten, Raymond N. and Vermeulen, Raymond F. L. and Twitchen, Daniel J. and Markham, Matthew and Hanson, Ronald},
journal = {Nature},
number = {7709},
pages = {268--273},
title = {{Deterministic delivery of remote entanglement on a quantum network}},
doi = {10.1038/s41586-018-0200-5},
volume = {558},
year = {2018}
}

@article{James2001,
author = {James, Daniel F. V. and Kwiat, Paul G. and Munro, William J. and White, Andrew G.},
journal = {Physical Review A},
number = {5},
pages = {052312},
title = {{Measurement of qubits}},
doi = {10.1103/PhysRevA.64.052312},
volume = {64},
year = {2001}
}

@article{Khabiboulline2019a,
author={Khabiboulline, Emil T. and Borregaard, Johannes and {De Greve}, Kristiaan and Lukin, Mikhail D.},
journal = {Physical Review Letters},
number = {7},
pages = {070504},
publisher = {American Physical Society},
title = {{Optical Interferometry with Quantum Networks}},
doi = {10.1103/PhysRevLett.123.070504},
volume = {123},
year = {2019}
}

@article{Khabiboulline2019b,
author={Khabiboulline, Emil T. and Borregaard, Johannes and {De Greve}, Kristiaan and Lukin, Mikhail D.},
journal = {Physical Review A},
number = {2},
pages = {022316},
publisher = {American Physical Society},
title = {{Quantum-assisted telescope arrays}},
doi = {10.1103/PhysRevA.100.022316},
volume = {100},
year = {2019}
}

@article{Kimble2008,
author = {Kimble, H. Jeff},
journal = {Nature},
number = {7198},
pages = {1023--1030},
title = {{The quantum internet}},
doi = {10.1038/nature07127},
volume = {453},
year = {2008}
}

@article{Komar2014,
author = {K{\'{o}}m{\'{a}}r, Peter and Kessler, Eric M. and Bishof, Michael and Jiang, Liang and S{\o}rensen, Anders S. and Ye, Jun and Lukin, Mikhail D.},
journal = {Nature Physics},
number = {8},
pages = {582--587},
title = {{A quantum network of clocks}},
doi = {10.1038/nphys3000},
volume = {10},
year = {2014}
}

@article{Lago-Rivera2021,
author = {Lago-Rivera, Dario and Grandi, Samuele and Rakonjac, Jelena V. and Seri, Alessandro and {de Riedmatten}, Hugues},
journal = {Nature},
number = {7861},
pages = {37--40},
publisher = {Springer US},
title = {{Telecom-heralded entanglement between multimode solid-state quantum memories}},
doi = {10.1038/s41586-021-03481-8},
volume = {594},
year = {2021}
}

@article{Lipinska2018,
author = {Lipinska, Victoria and Murta, Gl{\'{a}}ucia and Wehner, Stephanie},
journal = {Physical Review A},
number = {5},
pages = {052320},
publisher = {American Physical Society},
title = {{Anonymous transmission in a noisy quantum network using the W state}},
doi = {10.1103/PhysRevA.98.052320},
volume = {98},
year = {2018}
}

@article{Liu2021,
author = {Liu, Li-Zheng and Zhang, Yu-Zhe and Li, Zheng-Da and Zhang, Rui and Yin, Xu-Fei and Fei, Yue-Yang and Li, Li and Liu, Nai-Le and Xu, Feihu and Chen, Yu-Ao and Pan, Jian-Wei},
doi = {10.1038/s41566-020-00718-2},
journal = {Nature Photonics},
number = {2},
pages = {137--142},
title = {{Distributed quantum phase estimation with entangled photons}},
volume = {15},
year = {2021}
}

@article{Minar2008,
  title = {Phase-noise measurements in long-fiber interferometers for quantum-repeater applications},
  author = {Min{\'a}{\v{r}}, Ji{\v{r}}{\'\i} and {de Riedmatten}, Hugues and Simon, Christoph and Zbinden, Hugo and Gisin, Nicolas},
  journal = {Phys. Rev. A},
  volume = {77},
  issue = {5},
  pages = {052325},
  numpages = {8},
  year = {2008},
  publisher = {American Physical Society},
  doi = {10.1103/PhysRevA.77.052325}
}

@article{Monteiro2015,
  title = {Revealing Genuine Optical-Path Entanglement},
  author={Monteiro, Fernando and Vivoli, V. Caprara and Guerreiro, Thiago and Martin, Anthony and Bancal, Jean-Daniel and Zbinden, Hugo and Thew, Robert T. and Sangouard, Nicolas},
  journal = {Phys. Rev. Lett.},
  volume = {114},
  issue = {17},
  pages = {170504},
  numpages = {5},
  year = {2015},
  publisher = {American Physical Society},
  doi = {10.1103/PhysRevLett.114.170504}
}

@article{Morin2013,
  title = {Witnessing Trustworthy Single-Photon Entanglement with Local Homodyne Measurements},
  author = {Morin, Olivier and Bancal, Jean-Daniel and Ho, Melvyn and Sekatski, Pavel and D'Auria, Virginia and Gisin, Nicolas and Laurat, Julien and Sangouard, Nicolas},
  journal = {Phys. Rev. Lett.},
  volume = {110},
  issue = {13},
  pages = {130401},
  numpages = {5},
  year = {2013},
  publisher = {American Physical Society},
  doi = {10.1103/PhysRevLett.110.130401}
}

@phdthesis{Mosley2007,
title = {Generation of Heralded Single Photons in Pure Quantum States},
author = {Mosley, Peter J.},
year = {2007},
school = {University of Oxford},
url={https://ora.ox.ac.uk/objects/uuid:44c36e1e-11ee-41e2-ba29-611c932ce4ff}
}

@article{Murta2020,
author = {Murta, Gl{\'{a}}ucia and Grasselli, Federico and Kampermann, Hermann and Bru{\ss}, Dagmar},
journal = {Advanced Quantum Technologies},
number = {11},
pages = {2000025},
title = {{Quantum Conference Key Agreement: A Review}},
volume = {3},
year = {2020},
doi = {10.1002/qute.202000025}
}

@article{Papp2009,
author = {Papp, Scott B. and Choi, Kyung Soo and Deng, Hui and Lougovski, Pavel and van Enk, S. J. and Kimble, H. Jeff},
journal = {Science},
number = {5928},
pages = {764--768},
title = {{Characterization of Multipartite Entanglement for One Photon Shared Among Four Optical Modes}},
doi = {10.1126/science.1172260},
volume = {324},
year = {2009}
}

@article{Paris1996,
author = {Paris, Matteo G. A.},
journal = {Physics Letters A},
number = {2},
pages = {78--80},
title = {{Displacement operator by beam splitter}},
doi = {10.1016/0375-9601(96)00339-8},
volume = {217},
year = {1996}
}

@article{Sangouard2011,
author = {Sangouard, Nicolas and Simon, Christoph and {de Riedmatten}, Hugues and Gisin, Nicolas},
journal = {Reviews of Modern Physics},
number = {1},
pages = {33--80},
title = {{Quantum repeaters based on atomic ensembles and linear optics}},
doi = {10.1103/RevModPhys.83.33},
volume = {83},
year = {2011}
}

@article{Simon2007,
  title = {Quantum Repeaters with Photon Pair Sources and Multimode Memories},
  author = {Simon, Christoph and {de Riedmatten}, Hugues and Afzelius, Mikael and Sangouard, Nicolas and Zbinden, Hugo and Gisin, Nicolas},
  journal = {Phys. Rev. Lett.},
  volume = {98},
  issue = {19},
  pages = {190503},
  numpages = {4},
  year = {2007},
  publisher = {American Physical Society},
  doi = {10.1103/PhysRevLett.98.190503}
}

@article{Slodi2013,
  title = {Atom-Atom Entanglement by Single-Photon Detection},
  author = {Slodi{\v{c}}ka, L. and H{\'e}tet, G. and R{\"o}ck, N. and Schindler, P. and Hennrich, M. and Blatt, R.},
  journal = {Phys. Rev. Lett.},
  volume = {110},
  issue = {8},
  pages = {083603},
  numpages = {5},
  year = {2013},
  publisher = {American Physical Society},
  doi = {10.1103/PhysRevLett.110.083603}
}

@article{Stockill2017,
  title = {Phase-Tuned Entangled State Generation between Distant Spin Qubits},
  author={Stockill, Robert and Stanley, M. J. and Huthmacher, Lukas and Clarke, E. and Hugues, M. and Miller, A. J. and Matthiesen, C. and Le Gall, Claire and Atat{\"u}re, Mete},
  journal = {Phys. Rev. Lett.},
  volume = {119},
  issue = {1},
  pages = {010503},
  numpages = {6},
  year = {2017},
  publisher = {American Physical Society},
  doi = {10.1103/PhysRevLett.119.010503}
}

@article{Vivoli2015,
  title = {Comparing different approaches for generating random numbers device-independently using a photon pair source},
  author={Vivoli, Valentina Caprara and Sekatski, Pavel and Bancal, Jean-Daniel and Lim, Charles Ci Wen and Martin, Anthony and Thew, Robert T. and Zbinden, Hugo and Gisin, Nicolas and Sangouard, Nicolas},
  journal = {New Journal of Physics},
  volume = {17},
  number = {2},
  pages = {023023},
  year = 2015,
  publisher = {{IOP} Publishing},
	doi = {10.1088/1367-2630/17/2/023023}
}

@article{Wehner2018,
  title={Quantum internet: A vision for the road ahead},
  author={Wehner, Stephanie and Elkouss, David and Hanson, Ronald},
  journal={Science},
  volume={362},
  number={6412},
  pages={eaam9288},
  year={2018},
  publisher={American Association for the Advancement of Science},
  doi = {10.1126/science.aam9288}
}

@article{Yu2020,
author = {Yu, Yong and Ma, Fei and Luo, Xi-yu and Jing, Bo and Sun, Peng-fei and Fang, Ren-zhou and Yang, Chao-wei and Liu, Hui and Zheng, Ming-Yang and Xie, Xiu-Ping and Zhang, Wei-Jun and You, Li-Xing and Wang, Zhen and Chen, Teng-Yun and Zhang, Qiang and Bao, Xiao-Hui and Pan, Jian-Wei},
journal = {Nature},
number = {7794},
pages = {240--245},
publisher = {Springer US},
title = {{Entanglement of two quantum memories via fibres over dozens of kilometres}},
doi = {10.1038/s41586-020-1976-7},
volume = {578},
year = {2020}
}

@article{Zhong2018,
author = {Zhong, Han-Sen and Li, Yuan and Li, Wei and Peng, Li-Chao and Su, Zu-En and Hu, Yi and He, Yu-Ming and Ding, Xing and Zhang, W. -J. and Li, Hao and Zhang, Lu and Wang, Z. and You, L. -X. and Wang, Xi-Lin and Jiang, Xiao and Li, Li and Chen, Yu-Ao and Liu, Nai-Le and Lu, Chao-Yang and Pan, Jian-Wei},
journal = {Physical Review Letters},
number = {25},
pages = {250505},
title = {12-photon entanglement and scalable scattershot boson sampling with optimal entangled-photon pairs from parametric down-conversion},
doi = {10.1103/PhysRevLett.121.250505},
volume = {121},
year = {2018}
}
\end{document}